\documentclass[a4paper,11pt]{article}

\usepackage[height=8.85in,width=6.75in]{geometry}

\usepackage{chngcntr}
\counterwithin{equation}{section}

\usepackage{graphicx,rotating}
\usepackage[bookmarksopen,colorlinks=true,linkcolor=light_blue,citecolor=light_pink,urlcolor=dark_red,linktoc=all]{hyperref}
\usepackage{amsmath,mathtools,amsfonts,amssymb,slashed,braket,bm,bbm,cite,xcolor,bbold,physics,tensor,xfrac}
\usepackage{comment}
\usepackage[footnotesize]{caption}
\usepackage{fourier}

\newcommand{\Mpl}{M_{\text{Pl}}}
\def\thefootnote{\fnsymbol{footnote}}
\usepackage{multicol}
\definecolor{dark_red}{rgb}{0.7, 0., 0.}
\definecolor{light_pink}{rgb}{1,0.4,0.4}
\definecolor{light_blue}{rgb}{0.284602,0.317763,0.963947}
\definecolor{forestgreen}{HTML}{228B22}
\definecolor{ochre}{HTML}{CCAA2B}

\begin{document}
\hypersetup{pageanchor=false}
\begin{titlepage}

\begin{center}

\hfill KEK-TH-2545\\
\hfill KEK-Cosmo-0319\\
\hfill RESCEU-15/23\\
\vskip 0.3in
{\Huge \bfseries Quantum Corrections to\\
    Higgs Inflation in Einstein--Cartan Gravity\\}
\vskip .5in
{\Large
    Minxi He$^{a,b}$\footnote{
        {\tt heminxi``at''post.kek.jp}},
    Kohei Kamada$^{c,d,e}$\footnote{
        {\tt	kohei.kamada``at''ucas.ac.cn}
    },
    Kyohei Mukaida$^{a,f}$\footnote{
        {\tt	kyohei.mukaida``at''kek.jp
        }
    }
}

\vskip .2in
\begin{tabular}{ll}
$^a$& \!\!\!\!\!\emph{Theory Center, IPNS, KEK, } \\[-.3em]
& \!\!\!\!\!\emph{1-1 Oho, Tsukuba, Ibaraki 305-0801, Japan}\\
$^b$& \!\!\!\!\!\emph{Particle Theory and Cosmology Group, Center for Theoretical Physics of the Universe, }\\[-.3em] 
& \!\!\!\!\! \emph{Institute for Basic Science (IBS), 55, Expo-ro, Yuseong-gu, Daejeon, 34126, Korea}\\
$^c$& \!\!\!\!\!\emph{School of Fundamental Physics and Mathematical Sciences, Hangzhou Institute for Advanced Study, }
\\[-.3em]
& \!\!\!\!\! \emph{University of Chinese Academy of Sciences (HIAS-UCAS), Hangzhou 310024, China} \\
$^d$& \!\!\!\!\!\emph{Research Center for the Early Universe (RESCEU), Graduate School of Science,}
\\[-.3em]
& \!\!\!\!\! \emph{The University of Tokyo, Hongo 7-3-1, Bunkyo-ku, Tokyo 113-0033, Japan} \\
$^e$& \!\!\!\!\! \emph{International Centre for Theoretical Physics Asia-Pacific (ICTP-AP), Beijing/Hangzhou, China} \\
$^f$& \!\!\!\!\!\emph{Graduate Institute for Advanced Studies (Sokendai), }\\[-.3em]
& \!\!\!\!\!\emph{1-1 Oho, Tsukuba, Ibaraki 305-0801, Japan}\\
\end{tabular}

\end{center}

\vskip .2in

\begin{abstract}
\noindent
This paper studies the quantum corrections to the Higgs inflation model in the context of the Einstein--Cartan (E--C) gravity in the large-$ N $ limit with $N$ being the number of real scalar components in Higgs.
Recently, it is realized that the Higgs inflation in the E--C formalism smoothly connects those in the metric and the Palatini formalisms in the presence of a non-minimal coupling between the Higgs fields and the Nieh--Yan term.
This motivates us to investigate the quantum corrections in large-$ N $ limit to the E--C Higgs inflation and to clarify how the Ricci curvature squared $ R^2 $ induced by the quantum corrections succeeds in Ultraviolet (UV)-extending the Higgs inflation in metric formalism while it fails in the Palatini case.
We show that a generalized $ R^2 $-term required for the renormalization in the E--C formalism induces a new scalar degree of freedom (DoF), the scalaron, which gradually decouples with the system due to its increasing mass as approaching the Palatini limit.
The presence of the scalaron extends the UV cutoff at vacuum of the original model except for the parameter space close to the Palatini limit.
This UV-extension is expected to solve the strong coupling problem that may exist during (p)reheating in the absence of the scalaron.
\end{abstract}

\end{titlepage}

\tableofcontents
\thispagestyle{empty}
\renewcommand{\thepage}{\arabic{page}}
\renewcommand{\thefootnote}{$\natural$\arabic{footnote}}
\setcounter{footnote}{0}
\newpage
\hypersetup{pageanchor=true}

\renewcommand{\thefootnote}{$\sharp$\arabic{footnote}}
\setcounter{page}{1}
\setcounter{footnote}{0}
%%%%%%%%%%%%%%%%%%%%%%%%%%%%%%%%%%%%%%%%%%%%%%%%%%

%%%%%%%%%%%%%%%%%%%%%%%%%%%%%%%%%%%%%%%%%%%%%%%%%%
\section{Introduction}
\label{Sec-1}
%%%%%%%%%%%%%%%%%%%%%%%%%%%%%%%%%%%%%%%%%%%%%%%%%%

The leading paradigm of the early Universe is inflation~\cite{Starobinsky:1980te,Guth:1980zm,Sato:1980yn,Mukhanov:1981xt,Linde:1981mu,Albrecht:1982wi},\footnote{See also Ref.~\cite{Sato:2015dga} for a review.} which naturally solves several major problems in cosmology such as the large-scale homogeneity, isotropy, and flatness.
The origin of the inflaton field(s) which drives the exponential expansion of space during inflation remains unknown.
Various inflation models have been constructed and are constrained by the precise measurement of the Cosmic Microwave Background (CMB)~\cite{BENNETT1993409,Komatsu_2011,Planck:2018jri}.
The Higgs inflation (HI)~\cite{Cervantes-Cota:1995ehs,Bezrukov:2007ep} makes use of the only scalar field in the Standard Model (SM) of particle physics, the Higgs, as the role of inflaton with the help of a large non-minimal coupling $ \xi \gg 1 $ between Higgs and gravity.
This idea draws a wide range of attention because it gives observationally favored predictions on the scalar spectral index and tensor-to-scalar ratio and is regarded as a minimal setup.\footnote{See \textit{e.g.}, Ref.~\cite{Horn:2020wif} for a recent review of Higgs field in the early Universe.}

However, the viability of the predictions from HI is still under debate because it is found that HI enters a strongly-coupled regime during preheating~\cite{Ema:2016dny,Sfakianakis:2018lzf}.
This issue is of two folds. On the one hand, it is realized that the Ultraviolet (UV) cutoff of HI around the vacuum (in the metric formalism) is much lower than Planck scale $ \Lambda_{\rm metric} \sim \Mpl/\xi \ll \Mpl $~\cite{Burgess:2009ea,Barbon:2009ya,Burgess:2010zq,Hertzberg:2010dc,Barvinsky:2009ii,Bezrukov:2010jz}.
On the other hand, the typical momentum of the longitudinal mode of the gauge bosons excited during preheating $ \sim \sqrt{\lambda} \Mpl $ exceeds the cutoff $ \Lambda_{\rm metric} $, where $ \lambda $ is the self-coupling of the Higgs, which casts doubts on the predictability of results from perturbation theory.
A number of attempts to UV-extend HI have been made to resolve this problem~\cite{Giudice:2010ka,Barbon:2015fla,Ema:2017rqn,Lee:2018esk,Koshelev:2020fok}, among which the introduction of an $\alpha R^2 $ term with $ R $ being the Ricci scalar and $\alpha$ being a coupling is a natural choice because such a term is required to renormalize the theory~\cite{Salvio:2015kka,Netto:2015cba,Liu:2018hno}.
In addition, $\alpha R^2$ is the only term that becomes relevant below the Planck scale in the large $N$ analysis~\cite{Calmet:2016fsr,Ghilencea:2018rqg,Ema:2019fdd,Ema:2020zvg,Ema:2020evi}, with $ N $ being the number of real scalar fields in Higgs. The inherent non-perturbative nature of the large $N$ allows us to utilize this theory above $\Mpl / \sqrt{\alpha}$, which would imply a new dynamical scalar degree of freedom, \textit{i.e.}, the scalaron (see Sec.~\ref{Sec-3} for more detail).
The resulting mixed Higgs-$ R^2 $ inflation model, which smoothly connects the Starobinsky model~\cite{Starobinsky:1980te} and HI~\cite{Bezrukov:2007ep}, remains unitary up to Planck scale except for the parameter space close to the Higgs-inflation limit where the effective self-coupling of Higgs is larger than unity~\cite{Ema:2017rqn,Gorbunov:2018llf,He:2018mgb}.\footnote{The inflation dynamics~\cite{Wang:2017fuy,Ema:2017rqn,He:2018gyf,Gorbunov:2018llf,Gundhi:2018wyz,Enckell:2018uic} and reheating process~\cite{He:2018mgb,Bezrukov:2019ylq,He:2020ivk,Bezrukov:2020txg,He:2020qcb} of this two-field model have been intensively studied. Dark matter production~\cite{Aoki:2022dzd}, production of primordial black holes~\cite{Cheong:2019vzl,Cheong:2022gfc}, and magnetogenesis~\cite{Durrer:2022emo} in the mixed Higgs-$ R^2 $ model are also studied recently. See also Ref.~\cite{Cheong:2021vdb} for a review of recent progress. See Refs.~\cite{Canko:2019mud,Lee:2021dgi} for more general Higgs-scalaron model.}

Higgs inflation is not only considered in the metric formalism of gravity as in general relativity (GR) but also investigated in the Palatini case~\cite{Bauer:2010jg,Rasanen:2018ihz} which promotes the Levi-Civita connection uniquely determined by the spacetime metric to a general affine connection independent of the metric.
In this case, the UV cutoff around the vacuum shows a mild dependence on $\xi$ as
$ \Lambda_{\rm Palatini} \sim \Mpl/ \sqrt{\xi} $.
The (p)reheating process, on the other hand, shows no violent production of gauge bosons that violates the unitarity~\cite{Rubio:2019ypq}, contrary to the metric case.\footnote{See also Refs.~\cite{Gialamas:2019nly,Cheong:2021kyc,Lahanas:2022mng} for reheating process in inflation models in Palatini formalism.}
Similar to the metric case, $ R^2 $ term is also expected to renormalize the Palatini inflation model.
A single-field case with $ R^2 $ term is considered in Refs.~\cite{Enckell:2018hmo,Antoniadis:2018ywb}
which turns out to be equivalent to a specific $ P(\phi, X) $ theory (or $ k $-essence), instead of a two-field model with scalaron, with $ \phi $ a scalar field and $ X $ its kinetic term $ -\partial_{\mu} \phi \partial^{\mu} \phi /2 $.\footnote{Introduction of higher-dimensional operators in the non-minimal coupling and the potential in the Higgs inflation model in both metric and Palatini formalisms and the examination of their stability is shown in Ref.~\cite{Jinno:2019und}.}
The cutoff scale of this theory around the vacuum $ \phi=0 $ remains the same as $ \Lambda_{\rm Palatini} \sim \Mpl/ \sqrt{\xi} $
for a typical size of the coefficient of $ R^2 $ in the large-$N$ and -$\xi$ limit.
Generally speaking, the low cutoff scales in the Palatini models (and the metric ones as well) come from the non-trivial structure of the kinetic term or the non-trivial curvature of the four-dimensional target space of the Higgs fields (complex doublet),
since the Higgs potential is up to quartic power in the Jordan frame.
This motivates the attempt to UV-extend these models by embedding them into a flat space with one-dimension higher~\cite{Mikura:2021clt}. However, such an approach cannot UV-extend the Palatini case
up to the Planck scale
due to the presence of higher-dimensional operators in the potential after the embedding despite the flattened target space. Besides, a new model named the hybrid metric-Palatini Higgs inflation~\cite{He:2022xef} which smoothly connects the metric and the Palatini limits is proposed, aiming to further investigate the possibility of UV-extending the Palatini case. It was found that the embedding process considered can alleviate the situation but Planck-scale cutoff can only be realized in the metric limit.

Recently, HI is studied in the context of the Einstein--Cartan (E--C) gravity~\cite{Shaposhnikov:2020gts}.\footnote{A recent paper~\cite{Piani:2023aof} studies the oscillon formation during preheating in this model.} In addition to the non-minimal coupling $ \xi $ between Higgs and Ricci scalar, a new non-minimal coupling $ \xi_{\eta} $ between the Higgs and the Nieh--Yan term~\cite{Nieh:1981ww} is also introduced. Varying $ \xi_{\eta} $ from $ 0 $ to $ \xi $, the model smoothly goes from Palatini HI to metric HI, which shows that this model can be considered as a general model that includes metric and Palatini HI in a non-trivial way. Therefore, it is strongly motivated to investigate the UV property of this model for understanding the distinct behaviors of HI in different formalisms of gravity. Especially, if we consider the quantum corrections to this model in the large-$ N $ limit as done in the metric formalism~\cite{Ema:2019fdd}, it
receives non-trivial contribution from the Nieh--Yan term due to the non-minimal coupling $ \xi_{\eta} $, which enables us to systematically study how the UV-extended theories are related to each other and how the new DoF from $ R^2 $ (or scalaron) decouples with the system when taking the Palatini limit. Clarifying these questions may lead to a better understanding about the natural way to UV-extend HI in different formalisms, which is the main purpose of this paper.

This paper is constructed as follows. In Sec.~\ref{Sec-2}, the relevant basics of E--C gravity are introduced for the purpose of calculation in the rest of this paper, and a brief review of HI in E--C gravity
and its UV properties
are presented. Section~\ref{Sec-3} then shows the quantum corrections to the tree-level action in the large-$ N $ and -$\xi$ limit, which generally induces a new scalar DoF, the scalaron. Then, a discussion is presented on how the
scalaron behaves and
UV-extends the theory. Section~\ref{Sec-4} shows the effects of the scalaron on the dynamics during and after inflation with different choices of $ \xi_{\eta} $.
Summary and final conclusion are given in Sec.~\ref{Sec-5}.

%%%%%%%%%%%%%%%%%%%%%%%%%%%%%%%%%%%%%%%%%%%%%%%%%%
\section{Higgs Inflation in Einstein--Cartan Gravity}
\label{Sec-2}
%%%%%%%%%%%%%%%%%%%%%%%%%%%%%%%%%%%%%%%%%%%%%%%%%%

In this section, we will first introduce the relevant details of E--C gravity and its relation with the metric and the Palatini formalisms. Then, we review HI in E--C gravity with a non-minimally coupled Nieh--Yan term~\cite{Shaposhnikov:2020gts}, and show the metric and the Palatini HI models as two special limits in this model.

\subsection{Basics of Einstein--Cartan Gravity}
We adopt the sign convention $ (-1,1,1,1) $ for the spacetime metric $ g_{\mu\nu} $. The Levi-Civita connection associated with $ g_{\mu\nu} $ is uniquely defined as \begin{align}\label{eq-levi-civita}
    \Gamma\indices{^\rho_{\mu\nu}} \equiv \frac{1}{2} g^{\rho \sigma} \left( \partial_{\mu} g_{\nu\sigma} + \partial_{\nu} g_{\sigma \mu} - \partial_{\sigma} g_{\mu\nu}  \right) ~.
\end{align}
The covariant derivative associated with a general affine connection $ \bar{\Gamma} $ is defined as
\begin{align}
    \bar{\nabla}_{\mu} A\indices{^\nu_\rho} = \partial_{\mu} A\indices{^\nu_\rho} + \bar{\Gamma}\indices{^\nu_{\mu\lambda}} A\indices{^\lambda_\rho} - \bar{\Gamma}\indices{^\lambda_{\mu\rho}} A\indices{^\nu_\lambda}~.
\end{align}
The Riemann tensor, Ricci tensor, and Ricci scalar determined by $ \bar{\Gamma}\indices{^\rho_{\mu\nu}} $ are
\footnote{We follow the convention, for example, in Ref.~\cite{Carroll:2004st}.}
\begin{align}
    \bar{R}\indices{^\rho_{\sigma\mu\nu}} (\bar{\Gamma}) &\equiv \partial_{\mu} \bar{\Gamma}\indices{^\rho_{\nu\sigma}} - \partial_{\nu} \bar{\Gamma}\indices{^\rho_{\mu\sigma}} + \bar{\Gamma}\indices{^\rho_{\mu\lambda}} \bar{\Gamma}\indices{^\lambda_{\nu\sigma}} - \bar{\Gamma}\indices{^\rho_{\nu\lambda}} \bar{\Gamma}\indices{^\lambda_{\mu\sigma}} ~,\nonumber\\
    \bar{R}_{\mu\nu} (\bar{\Gamma}) &\equiv \bar{R}\indices{^\rho_{\mu\rho\nu}} (\bar{\Gamma}) ~,\nonumber\\
    \bar{R} (\bar{\Gamma}) &\equiv g^{\mu\nu} \bar{R}_{\mu\nu} (\bar{\Gamma}) ~.
\end{align}

The affine connection, $ \bar{\Gamma}\indices{^\rho_{\mu\nu}} $, is treated \textit{a priori} independent of the metric tensor in the E--C gravity whose dynamics is determined by its equation of motion, as a simple extension of GR.
This treatment is usually seen in the context of Palatini formalism or metric-affine formalism.\footnote{
    In the literature, Palatini formalism usually refers to the case where $ \Bar{\Gamma}\indices{^\rho_{\mu\nu}} = \Bar{\Gamma}\indices{^\rho_{\nu\mu}} $ and the matter sector does not directly couple to the $ \Bar{\Gamma}\indices{^\rho_{\mu\nu}} $, while there is no such restriction in the metric-affine case. $ f(R) $ theory has been intensively investigated in the Palatini and the metric-affine formalisms. See Refs.~\cite{Sotiriou:2006qn,Sotiriou:2008rp,Olmo:2011uz} for more detail.
    }
In principle, the two lower indices in $ \bar{\Gamma}\indices{^\rho_{\mu\nu}} $ are not necessarily symmetric, and the metric compatibility (\textit{i.e.,} metricity) can be violated.
The former implies the possible existence of torsion while the latter can lead to non-metricity.
In E--C gravity considered in this paper, the non-metricity is assumed to vanish while torsion is allowed.
On the other hand, in the Palatini gravity, torsion is usually absent but non-metricity exists.
As we will see shortly, we can impose the metricity condition without loss of generality in a certain class of theories.

Let us first clarify the relation between E--C gravity and Palatini gravity.
In a broad class of models with non-minimal coupling to gravity such as
\begin{align}
    \label{eq-nonminimal-fRphi}
    S_f= \int \dd^4x \sqrt{-g} \left[ f\left( \bar{R},\phi \right) -\frac{1}{2} g^{\mu\nu} \partial_{\mu} \phi \partial_{\nu} \phi \right]
\end{align}
with $ \phi $ being some scalar field,
we can show that these two formalisms are connected by the so-called projective transformation:
\begin{align}\label{eq-projective-transformation}
    g_{\mu\nu} \to g_{\mu\nu} ~, \qquad
    \bar{\Gamma}\indices{^\rho_{\mu\nu}} \to \bar{\Gamma}\indices{^\rho_{\mu\nu}} +\delta_{\nu}^{\rho} U_{\mu} ~,
\end{align}
where $ U_{\mu} $ is an arbitrary vector field.
Since $\bar{R} (\bar{\Gamma}) $ is invariant under \eqref{eq-projective-transformation}, the action \eqref{eq-nonminimal-fRphi} respects the projective symmetry.
As this is nothing but the redundancy of the description, a particular choice of $U_\mu$ can be regarded as a gauge choice~\cite{Hehl:1994ue,Julia:1998ys,Dadhich:2012htv}.\footnote{
    In general, if the action is constructed with Ricci scalar and there is no direct coupling between matter and $ \bar{\Gamma}\indices{^\rho_{\mu\nu}} $, different gauge choices give equivalent theories. (If the matter sector does not satisfy the projective symmetry, it may cause inconsistency~\cite{Sotiriou:2006qn}.)
    If we further restrict the action to contains only the Ricci scalar, a class of theories related through a particular choice of $U_\mu$ are equivalent to GR.}
To show the equivalence and how a particular choice of $ U_{\mu} $ (which obviously has fewer DoFs to eliminate the full torsion tensor in Eq.~\eqref{eq-def-torsion}) connects seemingly different theories, we decompose the affine connection as
\begin{align}\label{eq-decomposition-affine-connection}
    \bar{\Gamma}\indices{^\rho_{\mu\nu}} = \Gamma\indices{^\rho_{\mu\nu}} + C\indices{^\rho_{\mu\nu}} ~,
\end{align}
where the deviation from the Levi-Civita connection of the metric is encoded in the new tensor $ C\indices{^\rho_{\mu\nu}} $.
If we insert Eq.~\eqref{eq-decomposition-affine-connection} into Eq.~\eqref{eq-nonminimal-fRphi}, the solution of the constraint equation for $ C\indices{^\rho_{\mu\nu}} $ shows
\begin{align}\label{eq-solution-C}
     C\indices{^\rho_{\mu\nu}}= \frac{1}{2} \left( \delta_{\mu}^{\rho} \partial_{\nu} \ln \left| \frac{\partial f}{\partial \bar{R}} \right| - g^{\rho\sigma} g_{\mu\nu} \partial_{\sigma} \ln \left| \frac{\partial f}{\partial \bar{R}} \right| \right) + \frac{1}{2} \delta_{\nu}^{\rho} \left( \partial_{\mu} \ln \left| \frac{\partial f}{\partial \bar{R}} \right| -\frac{2}{3} T\indices{^\sigma_{\sigma\mu}} \right) ~,
\end{align}
where the torsion tensor is defined as
\begin{align}\label{eq-def-torsion}
    T\indices{^\rho_{\mu\nu}} \equiv \bar{\Gamma}\indices{^\rho_{\mu\nu}} - \bar{\Gamma}\indices{^\rho_{\nu\mu}} =C\indices{^\rho_{\mu\nu}} -C\indices{^\rho_{\nu\mu}} = - T\indices{^\rho_{\nu\mu}} ~,
\end{align}
which represents the antisymmetric part of the affine connection.
Only the trace of $ T\indices{^\rho_{\mu\nu}} $ enters Eq.~\eqref{eq-solution-C} and it cannot be fixed by the constraint equation, which implies the existence of gauge DoFs that can be sufficiently controlled by a vector field [\textit{i.e.,} $ U_{\mu} $ in Eq.~\eqref{eq-projective-transformation}]. One can easily check that non-metricity appears but torsion vanishes if $ T\indices{^\sigma_{\sigma\mu}} =0 $, while torsion appears with metricity condition satisfied if
the last parentheses in Eq.~\eqref{eq-solution-C} is zero.
Hence, the Palatini and E--C formalisms are equivalent up to a gauge choice of $U_\mu$. If we further assume the absence of non-minimal coupling and higher-order operators in $ \bar{R} $, $ \partial f/ \partial \bar{R} ={\rm const.} $ such that the theory reproduces GR up to a non-dynamical $ T\indices{^\sigma_{\sigma\mu}} $. Therefore, the Palatini HI~\cite{Bauer:2010jg}, whose action is given in the form of Eq.~\eqref{eq-nonminimal-fRphi},  can be equivalently described in the context of E--C gravity.

For the purpose of this paper, it is necessary to take into account a (non-minimal coupled) Nieh--Yan term which allows us to smoothly extend the Palatini HI to a broader theory that includes the metric HI, as proposed in Ref.~\cite{Shaposhnikov:2020gts}.
The (non-minimal coupled) Nieh--Yan term [see Eq.~\eqref{eq-j-action-NY}] can also be shown to be invariant under the projective transformation
\begin{align}\label{eq-projective-transform-NY}
    E^{\mu\nu\rho\sigma} g_{\nu\alpha} T\indices{^\alpha_{\rho\sigma}} \to & ~ E^{\mu\nu\rho\sigma} g_{\nu\alpha} \left( T\indices{^\alpha_{\rho\sigma}} + \delta_{\sigma}^{\alpha} U_{\rho} - \delta_{\rho}^{\alpha} U_{\sigma} \right) \nonumber \\
    &= E^{\mu\nu\rho\sigma} g_{\nu\alpha} T\indices{^\alpha_{\rho\sigma}} + E^{\mu\nu\rho\sigma} \left( g_{\nu\sigma} U_{\rho} - g_{\nu\rho} U_{\sigma} \right) ~,
\end{align}
where the last part vanishes because $ g_{\mu\nu} = g_{\nu\mu} $ and
$ E^{\mu\nu\rho\sigma} $ is the totally anti-symmetric tensor with $E^{\mu\nu\rho\sigma} = \epsilon^{\mu\nu\rho\sigma} / \sqrt{-g}$ and $\epsilon^{0123} = 1$.
However, the torsion can no longer be removed solely by the choice of a vector field $U_\mu$ in the presence of such a term because it contains additional DoFs from $ T\indices{^\rho_{\mu\nu}} $ as seen in Eq.~\eqref{eq-projective-transform-NY}. Consequently, torsion cannot be eliminated only by $ T\indices{^\sigma_{\sigma\mu}}=0 $. Still, the metricity condition can be achieved by properly choosing the gauge $U_\mu$.
Therefore, we can take the E--C formalism, \textit{i.e.,} vanishing non-metricity, without loss of generality.

In order to clarify the relation between the metric formalism and E--C formalism, we separate the contributions from $ g_{\mu\nu} $ and $ \bar{\Gamma}\indices{^\rho_{\mu\nu}} $ in the following calculation.
Owing to the vanishing non-metricity in E--C formalism, the deviation from the Levi-Civita connection $C\indices{^\rho_{\mu\nu}}$ can be expressed by the torsion tensor as
\begin{align}
    C\indices{^\rho_{\mu\nu}} = \frac{1}{2} \qty( T\indices{^\rho_{\mu\nu}} + T\indices{_\nu^\rho_\mu} - T\indices{_{\mu\nu}^\rho} ),
\end{align}
which is nothing but the contorsion tensor.
By utilizing this fact, we can decompose the Ricci scalar as follows~\cite{Shapiro:2014kma}\footnote{There is a sign difference from ref.~\cite{Shapiro:2014kma} merely due to different conventions of covariant derivative.}
\begin{align}\label{eq-decomposed-R}
    \bar{R} \equiv g^{\mu\nu}\bar{R}_{\mu\nu}(\bar{\Gamma}) = R + 2 \nabla_{\mu} T^{\mu} - \frac{2}{3} T_{\mu} T^{\mu} + \frac{1}{24} S_{\mu} S^{\mu} + \frac{1}{2} q^{\mu\nu\rho} q_{\mu\nu\rho} ~,
\end{align}
where $ R \equiv \bar{R} (\Gamma) $ is uniquely determined by $ g_{\mu\nu} $, and
\begin{alignat}{2}
    &\text{(vector)}& \quad
    T_{\mu} &\equiv T\indices{^\alpha_{\mu\alpha}}
    ~,\nonumber\\
    &\text{(axial vector)}& \quad
    S^{\beta} &\equiv E^{\mu\nu\alpha\beta} T_{\mu\nu\alpha}
    ~,\nonumber\\
    &\text{(tensor)}& \quad
    q_{\alpha\beta\gamma} & \equiv T_{\alpha\beta\gamma} - \frac{1}{3} \left( g_{\alpha\gamma} T_{\beta}- g_{\alpha\beta} T_{\gamma} \right) +\frac{1}{6} E_{\alpha\beta\gamma\mu} S^{\mu} ~.
    \label{eq-decomposition-T}
\end{alignat}
One can show that
$ q_{\alpha\beta\gamma} $ is traceless $q\indices{^\alpha_{\mu\alpha}} =0$ and satisfies the Bianchi-type identity $ E^{\mu\nu\alpha\beta} q_{\mu\nu\alpha} =0 $.
The covariant derivative $ \nabla $ on the right hand side of Eq.~\eqref{eq-decomposed-R} is associated with the Levi-Civita connection of $ g_{\mu\nu} $.
While the vector and the axial vector couple to the Higgs in a non-trivial way in this paper, the action of the tensor part is always quadratic by itself $\propto q_{\alpha \beta \gamma}^2$ and hence can be integrated out trivially.
For this reason, we omit the tensor part $q_{\alpha \beta \gamma}$ in the following discussion for notational brevity.

\subsection{Higgs Inflation}
The action of HI in E--C gravity with non-minimally coupled Nieh--Yan term is given in the Jordan frame as~\cite{Shaposhnikov:2020gts}\footnote{The Holst term is also considered in Ref.~\cite{Shaposhnikov:2020gts} but it is irrelevant in the current discussion.}
\begin{align}\label{eq-total-action-jordan}
    S= S_{\rm grav, J}+ S_{\text{N--Y}, J} +S_{\rm Higgs, J} ~,
\end{align}
with
\begin{align}
	S_{\rm grav, J} &= \int \dd^4x \sqrt{-g_{\rm J}} F(\phi) g_{\rm J}^{\mu \nu} \bar{R}_{\mu\nu} (\bar{\Gamma}) ~, \label{eq-j-action-grav} \\
	S_{\text{N--Y}, J}&= -\frac{\xi_{\eta}}{4} \int \dd^4x ~ G(\phi) \partial_{\mu} \left( \sqrt{-g_{\rm J}} E^{\mu\nu\rho\sigma} T_{\nu\rho\sigma} \right) ~, \label{eq-j-action-NY} \\
	S_{\rm Higgs, J}&= \int \dd^4x \sqrt{-g_{\rm J}} \left[ -\frac{1}{2} g_{\rm J}^{\mu\nu} \partial_{\mu} \phi^i \partial_{\nu} \phi_i - V(\phi) \right] ~,
    \label{eq-j-action-Higgs}
\end{align}
where $ \phi^i $ are real scalar fields with $ i= 1,2, \cdots, N $ and $ \phi^2 \equiv \delta_{ij} \phi^i \phi^j $.
The subscript ``N--Y'' represents Nieh--Yan.
In the case of SM Higgs, $ N=4 $.
$ V(\phi) $, $ F(\phi) $, and $ G(\phi) $ are three arbitrary functions of $ \phi \equiv \sqrt{\phi^2} $. For the specific case considered in this paper, we take
\begin{align}
    \label{eq-FandG-func}
    V(\phi) = \Lambda + \frac{1}{2} m^2 \phi^2 + \frac{1}{4} \lambda \phi^4 ~, \qquad
    F(\phi) = \frac{\Mpl^2}{2} \left( 1+ \xi \frac{\phi^2 }{\Mpl^2} \right)~, \qquad
    G(\phi) = \phi^2 ~,
\end{align}
where we denote a cosmological constant as $\Lambda$,
the Higgs mass as $m$,
the Higgs four-point coupling as $\lambda$,
the reduced Planck scale as $ \Mpl =1/\sqrt{8\pi G_N} $ with $ G_N $ the gravitational constant, and the non-minimal coupling between Higgs and gravity as $ \xi $.\footnote{The conformal coupling corresponds to $ \xi =-1/6 $.} We mainly focus on the regime $ \xi \gg 1 $ and $ |\xi_{\eta}| \lesssim \xi $, because $ |\xi_{\eta}| \gtrsim \xi $ can easily violate the observational bound~\cite{Planck:2018jri} on the tensor-to-scalar ratio~\cite{Shaposhnikov:2020gts}. It will become evident soon that there is a discrete symmetry $ \xi_{\eta} \to -\xi_{\eta} $ in this system without other matter sectors, so we will only discuss positive $ \xi_{\eta} $ hereafter. Given such a matter sector, one can easily find that $ q_{\alpha\beta\gamma} =0 $. And $ S^{\mu} $ vanishes if we further remove the Nieh--Yan term by setting $ \xi_{\eta} =0 $.

The action~\eqref{eq-total-action-jordan} is invariant under the projective transformation~\eqref{eq-projective-transformation}. This model smoothly connects HI models in the metric and the Palatini formalisms~\cite{Shaposhnikov:2020gts}.
In the limit $ \xi_{\eta} =0 $, it is obviously the same as Palatini HI because they differ only by a gauge transformation as we have shown.
In the limit $ \xi_{\eta} =\xi $, it is not so straightforward to see the coincidence with the metric formalism. One can solve the constraint equations for $ \bar{\Gamma}\indices{^\rho_{\mu\nu}} $ to go back to the corresponding metric theory, which shows that the non-minimally coupled Nieh--Yan term basically provides an additional contribution to the non-canonical kinetic term of $ \phi^i $ to compensate the difference between the Palatini HI and the metric case.
We will explicitly show that at the end of Sec.~\ref{sec-conformal-mode}.

\subsection{Conformal Mode}\label{sec-conformal-mode}

To see how HI in E--C formalism given in Eqs.~\eqref{eq-total-action-jordan}--\eqref{eq-FandG-func} smoothly connects the metric and Palatini formalisms, we expand the action by using the decompositions in Eqs.~\eqref{eq-decomposed-R} and \eqref{eq-decomposition-T}.
Furthermore, the action enjoys the redundancy of the description, which is associated with the Weyl transformation.
In the following, we extract the conformal mode of the metric following Ref.~\cite{Ema:2020zvg}, \textit{e.g.}, in the Jordan frame,
\begin{align}\label{eq-jordan-tilde}
    g_{{\rm J}\mu\nu} = \frac{\Phi_{\rm J}^2}{6 \Mpl^2} \tilde{g}_{\mu\nu} ~ , \qquad \tilde{g} \equiv \det \left[ \tilde{g}_{\mu\nu} \right] =-1 ~.
\end{align}
Now the Weyl transformation between different frames becomes a field redefinition involving $\Phi_{\text{J}}$ and $\phi_i$, and the physical quantities should be invariant under such redefinitions.
The choice of a ``frame'' refers the choice of the conformal mode.\footnote{
    See Refs.~\cite{Ema:2020zvg,Ema:2020evi} for more details.
    }
We also redefine the following quantities
\begin{align}\label{eq-redefine-Jordan}
    \tilde{\phi}^i \equiv \frac{\Phi_{\rm J}}{\sqrt{6} \Mpl} \phi^i~, \qquad
    \tilde{T}_\mu = T_\mu~, \qquad
    \tilde{T}^\mu = \tilde g^{\mu\nu} \tilde T_\nu~, \qquad
    \tilde{S}^{\mu} \equiv \epsilon^{\mu\nu\rho\sigma} \tilde g_{\nu \alpha} T\indices{^\alpha_{ \rho\sigma}} =
    \frac{\Phi_{\rm J}^2}{6\Mpl^2} S^{\mu} ~, \qquad
    \tilde{S}_{\mu} = \tilde{g}_{\mu\nu} \tilde{S}^{\nu} ~.
\end{align}
By utilizing Eqs.~\eqref{eq-jordan-tilde} and \eqref{eq-redefine-Jordan}, we can rewrite the action given in Eqs.~\eqref{eq-total-action-jordan}--\eqref{eq-FandG-func} as
\begin{align}
    \label{eq-Jordan-conformal-mode-1}
    S= \int \dd^4 x &
    \Bigg[
        \frac{\Phi_{\rm J}^2}{12} \tilde{R} +\frac{1}{2} \tilde{g}^{\mu\nu} \left( \partial_{\mu} \Phi_{\rm J} \partial_{\nu} \Phi_{\rm J} -\partial_{\mu} \tilde{\phi}^i \partial_{\nu} \tilde{\phi_i} \right)
        - \Bigg( \frac{\Phi_{\rm J}^2}{6\Mpl^2} \Bigg)^2 V(\phi)
        - \frac{1}{2} \tilde \phi^2 \frac{\tilde \Box \Phi_{\text{J}}}{\Phi_{\text{J}}}
        - \frac{\Phi_{\rm J}^2}{18} \left( \tilde{T}^2 - \frac{1}{16} \tilde{S}^2 \right)
    \Bigg] \nonumber\\
    & + \int \dd^4 x~ \frac{\xi}{2} \tilde{\phi}^2 \left(\tilde{R} + A_{R} + r A_\text{N--Y} \right) ~,
\end{align}
where
\begin{align}
    \label{eq-def-A}
    A_{R} &\equiv - \frac{2}{3} \left(
        \tilde{T}^2 - \frac{1}{16} \tilde{S}^2
        \right)
    + 2\tilde{\nabla}_{\mu} \tilde{T}^{\mu}+4 \tilde T^{\mu} \partial_{\mu} \ln \left| \Phi_{\rm J} \right| - 6 \frac{\tilde{\Box}\Phi_{\rm J}}{\Phi_{\rm J}} ~, \qquad
    A_\text{N--Y} \equiv \frac{1}{2} \tilde{\nabla}_{\mu} \tilde{S}^{\mu} + \tilde{S}^{\mu} \partial_{\mu} \ln \left| \Phi_{\rm J} \right| ~, \qquad
    r \equiv \frac{\xi_\eta}{\xi}~.
\end{align}

Since the action \eqref{eq-jordan-tilde} is quadratic in $\tilde T_\mu$ and $\tilde S_\mu$, one may readily solve the constraint equations by completing the squares, which reads
\begin{align}
    \label{eq-jordan-tilde2}
    S= & \int \dd^4 x
    \Bigg[
        \frac{\Phi_{\rm J}^2}{12} \Omega_0^2 \tilde{R}
        +\frac{1}{2} \tilde{g}^{\mu\nu}
        \left(
            \partial_{\mu} \Phi_{\rm J} \partial_{\nu} \Phi_{\rm J} -\partial_{\mu} \tilde{\phi}^i \partial_{\nu} \tilde{\phi_i}
        \right)
        - \left(
            \frac{\Phi_{\rm J}^2}{6\Mpl^2}
        \right)^2 V(\phi)
        - \qty( 6 \xi + 1 ) \frac{\tilde{\phi}^2}{2} \frac{\tilde{\Box}\Phi_{\rm J}}{\Phi_{\rm J}}
    \Bigg]  \nonumber\\
    &+ \int \dd^4 x \left[ \left( 1-r^2 \right) \frac{\Phi_{\rm J}^2}{8} \Omega_0^2 \tilde{g}^{\mu\nu} \partial_{\mu} \ln \left| \Omega^2_0 \right| \partial_{\nu} \ln \left| \Omega^2_0 \right| \right] \nonumber \\
    &- \int \dd^4 x~ \frac{\Phi_{\rm J}^2}{18} \Omega^2_0 \left[ \left( \tilde T_{\mu} + \frac{3}{2} \partial_{\mu} \ln \left| \Omega^2_0 \right| \right)^2 - \frac{1}{16} \left( \tilde{S}_{\mu} - 6r \partial_{\mu} \ln \left| \Omega^2_0 \right| \right)^2 \right] ~.
\end{align}
We define the conformal factor as
\begin{align}
    \label{eq-Omega-tree}
    \Omega^2_0 \equiv 1 + 6 \xi \frac{\tilde \phi^2}{\Phi_{\text{J}}^2}~,
\end{align}
where the subscript ``$0$'' represents the result without quantum corrections.
It is clear that the third line provides the solution to $\tilde T_\mu$ and $\tilde S_\mu$ while being decoupled from the rest.
As we will see shortly in Eq.~\eqref{eq-Einstein-action-tilde1},
the scalar fields and the conformal mode form a \textit{nonlinear sigma model} whose target space is curved.
This implies the existence of a UV cutoff scale besides the Planck scale.

Let us confirm that this action reproduces HI in the metric and Palatini formalisms in the limit of $r = 1$ and $r = 0$ respectively.
For $r = 1$, the second line in Eq.~\eqref{eq-jordan-tilde2} vanishes.
By noting that
\begin{align}
    - \int \dd^4 x~ \qty( 6 \xi + 1 ) \frac{\tilde \phi^2}{2} \frac{\tilde \Box \Phi_{\text{J}}}{\Phi_{\text{J}}}
    =
    \int \dd^4 x~ \qty( 6 \xi + 1 ) \qty(
        \frac{\tilde\phi^i}{\Phi_{\text{J}}} \tilde g^{\mu\nu} \partial_\mu \tilde \phi_i \partial_\nu \Phi_{\text{J}}
        -
        \frac{1}{2} \frac{\tilde \phi^2}{\Phi_{\text{J}}^2} \tilde g^{\mu\nu} \partial_\mu \Phi_{\text{J}} \partial_\nu \Phi_{\text{J}}
    )~,
\end{align}
one can see that the first line in Eq.~\eqref{eq-jordan-tilde2} coincides with Eq.~(2.7) in Ref.~\cite{Ema:2020zvg}, which is nothing but the metric HI.
On the other hand, for $r = 0$, the E--C and Palatini formalisms are equivalent up to the redundancy of the projective transformation as we have shown.
We can also confirm this equivalence explicitly.
For instance, solving the constraint equation of $\tilde T_\mu$, one finds the Ricci curvature \eqref{eq-decomposed-R}
\begin{align}
    \bar R =
    R - \frac{3}{2} g^{\mu\nu} \partial_\mu\ln \left| \Omega^2_0 \right| \partial_\nu \ln \left| \Omega^2_0 \right| - 3 \Box \ln \left| \Omega^2_0 \right|~,
\end{align}
which coincides with the known result in the Palatini formalism (see \textit{e.g.,} Ref.~\cite{Shaposhnikov:2020gts}).

\subsection{Cutoff Scale and Perturbativity}
\label{sec-cutoff-tree}

As is known in the literature, HI in both metric and Palatini formalisms have cutoff scales much smaller than the Planck scale, \textit{i.e.}, $\Lambda_\text{metric} \sim \Mpl / \xi $ and $\Lambda_\text{Palatini} \sim \Mpl / \sqrt{\xi}$, respectively.
Here we estimate the cutoff scale for general $r$ and see how the metric and Palatini HI are smoothly connected by changing $r$ from $1$ to $0$.

Before discussing the technical detail, we clarify the definition of ``cutoff scale'' or ``unitarity violation scale'' adopted in this paper. We can estimate the cutoff scale of a theory in several ways, such as by the (tree-level) scattering amplitude exceeding unity or the higher-order contributions exceeding the lower-order ones in perturbative expansion. In the current paper, we adopt the former as in some other papers (see Ref.~\cite{Hertzberg:2010dc} for example) to define the cutoff scale beyond which the system enters the strongly coupled regime. Specifically, we require the tree-level four-point scattering amplitude of Higgs to be smaller than unity to maintain perturbativity. Therefore, in this section, we determine the cutoff scale of the system~\eqref{eq-jordan-tilde2} by calculating the tree-level four-point scattering of Higgs which receives contributions from the non-canonical kinetic terms and the potential.

The four-point amplitudes induced by the kinetic terms are known to be proportional to the Riemann curvature tensor of the target space, (see related discussion in Refs.~\cite{Alonso:2015fsp,Alonso:2016oah,Nagai:2019tgi,Cohen:2021ucp,Cheung:2021yog,Helset:2022tlf}), which is covariant under field redefinition (or equivalently, coordinate transformation)~\cite{Cohen:2021ucp}.
As we extract the conformal mode of the metric, the coordinate transformation also involves the Weyl transformation, thereby the amplitudes are in general frame-covariant.
Moreover, as we will show later, the scattering amplitudes at vacuum are accidentally frame-independent.
Therefore, one can use any field basis for convenience.
In our case, it is convenient to take the conformal mode in the Einstein frame $\Phi_\text{E}$ because the calculation of the Riemann tensor becomes simpler.
In the form of Eq.~\eqref{eq-jordan-tilde2} which corresponds to the Jordan frame, transforming into the Einstein frame simply means changing the conformal mode $ \Phi_{\rm J} $ to the one in the Einstein frame
\begin{align}\label{eq-Einstein-metric-0}
    \Phi_{\rm E} \equiv \Omega_0 \Phi_{\rm J} ~,
    \qquad
    g_{{\rm E}\mu\nu} = \frac{\Phi_{\rm E}^2}{6\Mpl^2} \tilde{g}_{\mu\nu} ~.
\end{align}
Note that, generally speaking, we can also redefine the Higgs field in the Einstein frame, analogous to that in Eqs.~\eqref{eq-redefine-Jordan}. However, we are not going to do so because we found it simpler by using $ \phi_i $.
The resulting action becomes
\begin{align} \label{eq-Einstein-action-tilde1}
    S= \int \dd^4 x &
    \Bigg[
        \frac{\Phi_{\rm E}^2}{12} \tilde{R}
        +\frac{1}{2} \tilde{g}^{\mu\nu}
        \left(
            \partial_{\mu} \Phi_{\rm E} \partial_{\nu} \Phi_{\rm E} - \frac{\Phi_\text{E}^2/\Omega_0^2}{6\Mpl^2} \partial_{\mu} \phi_i \partial_{\nu} \phi^i
        \right)
        - \left( \frac{\Phi_\text{E}^2/\Omega_0^2}{6\Mpl^2} \right)^2 V(\phi) \nonumber \\
    &- r^2 \frac{\Phi_\text{E}^2}{8} \tilde{g}^{\mu\nu} \partial_{\mu} \ln \left| \Omega_0^2 \right| \partial_{\nu} \ln \left| \Omega_0^2 \right|
    \Bigg] ~.
\end{align}
By rewriting this action in terms of $ g_{\text{E}\mu\nu} $ given in Eq.~\eqref{eq-Einstein-metric-0}, we can recover the usual Einstein--Hilbert term as shown later in Eq.~\eqref{eq-EC-tree-usual-Einstein}.

The unavoidable UV cutoff of the effective theory arises when the graviton becomes strongly coupled.
In Eq.~\eqref{eq-Einstein-action-tilde1}, the VEV of the conformal mode, $ \Phi_\text{E}$, characterizes the normalization of the graviton kinetic term instead of $ \Mpl $, which is simply because we have rescaled the metric as in Eq.~\eqref{eq-Einstein-metric-0} which measures ``distance" in the spacetime.
Hence, the graviton loops can be safely neglected if we restrict ourselves to $\tilde\ell  \Phi_\text{E}/\sqrt{6} > 1$ with $\tilde\ell$ being a length scale measured by $\tilde g_{\mu\nu}$. This requirement is equivalent to $\ell_\text{E} \Mpl > 1$ with $\ell_\text{E}$ being measured by $g_{\text{E}\mu\nu}$ [see Eq.~\eqref{eq-Einstein-metric-0}].
In this way, dimensional quantities defined on $\tilde g_{\mu\nu}$ should be compared to the unavoidable UV cutoff $\Phi_\text{E}$, equivalent to those on $g_{\text{E}\mu\nu}$ to $\Mpl$, which is usually used in the literature.

The obtained action \eqref{eq-Einstein-action-tilde1} involves a curved target space
as $ \Omega_0 $ is a function of $ \phi^i $.
The target-space metric in a field basis of $(\Phi_\text{E}, \phi^i)$ reads
\begin{align}
    \dd s^2 & = - \dd \Phi_{\rm E}^2 +\frac{1}{6} \frac{\Phi_{\rm E}^2}{\Mpl^2+\xi \phi^2}
        \left(
            \delta_{ij} + \frac{6r^2\xi^2}{\Mpl^2+\xi \phi^2}  \phi_i \phi_j
        \right) \dd\phi^i \dd\phi^j ~.
\end{align}
We are interested in the cutoff scale at vacuum, so we calculate the Riemann tensor at $ (\bar{\Phi}_{\rm E}, \bar{\phi}^i)=(\sqrt{6} \Mpl, \Vec{0}) $ whose non-vanishing components  are given as\footnote{
    In the metric limit, $|r| = 1$, the field-space curvature becomes flat only if $\xi = -1/6$ as the Higgs is decoupled from the conformal mode, \textit{i.e.,} conformal coupling, as indicated by the Planck unit.
}
\begin{align}\label{eq-riemann-HI}
    \bar{R}_{ikjl}\equiv R_{ikjl} (\bar{\Phi}_{\rm E},\bar{\phi}^i) &= R_N \left( \delta_{ij} \delta_{kl} - \delta_{il} \delta_{kj} \right) ~, \quad R_N \equiv \frac{1+12\xi (1+3r^2\xi)}{6\Mpl^2} ~.
\end{align}
Here we use the bar to indicate that the fields are in the vacuum (but not related to the general affine connection).
One immediately notices that all the components involving $ \Phi_{\rm E} $ vanish, so there is no mixing between $ \Phi_{\rm E} $ and $ \phi^i $ through the Riemann tensor,
and hence no scattering process such as $\Phi_\mathrm{E} \Phi_\mathrm{E} \rightarrow \phi_i \phi_i$ occurs.
This is also an advantage of the formalism utilizing the field space curvature,
otherwise we need to confirm the cancellation of the diagrams for the scattering of $\Phi_\mathrm{E}$.
Note that we choose $ \bar{\Phi}_{\rm E} = \sqrt{6}\Mpl $ at vacuum because it is a conventional choice to take the Planck scale as the ultimate UV cutoff where the graviton loops cannot be neglected [see also Eq.~\eqref{eq-EC-tree-usual-Einstein}].\footnote{
    One might wonder how come the results are independent of the VEV of $\Phi_\text{E}$.
    As explained already, what is truly physical is the ratio between the dimensional quantity and the cutoff scale where the graviton becomes strongly coupled.
    See Appendix~\ref{app-PhiE-vev} for more details.
}
Interestingly, the Riemann tensor at vacuum is \textit{frame-independent}.
As an illustration, consider a field basis corresponding to the Jordan frame, \textit{i.e.}, $\Phi_\text{E}^2 = \Phi_\text{J}^2 \qty( 1 + \xi \phi_i^2 / \Mpl^2 )$.
To transform the Riemann tensor \eqref{eq-riemann-HI} at vacuum, what we need is a field transformation at vacuum, which is trivial, \textit{i.e.,} $\qty(\partial \Phi_\text{E} / \partial \Phi_J)_\text{vac} = 1$.
Hence the Riemann tensor at vacuum is frame-independent.

\begin{figure}[t]
	\centering
 	\includegraphics[width=0.2\linewidth]{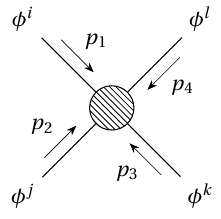}
	\caption{
        Diagram corresponding to the four-point scattering amplitude given in Eq.~\eqref{eq-fourpoint}.
}
	\label{fig:fourpoint}
\end{figure}

Consequently, one can estimate the cutoff scale for the E--C HI without quantum corrections by considering the scattering process $ \phi^i \phi^i \to \phi^j \phi^j $ whose interaction is charaterized by Eq.~\eqref{eq-riemann-HI}.
The results in Refs.~\cite{Nagai:2019tgi,Cohen:2021ucp}
show that the kinetic term contribution to tree-level four-point amplitudes at $ (\bar{\Phi}_\text{E},\bar{\phi}^i) $ can be written in a covariant form
\begin{align} \label{eq-fourpoint}
     \left. \mathcal{M}_{\phi^i \phi^j \to \phi^k \phi^l} \right|_\text{kin} = \frac{2}{3} \left[ s_{12} \bar{R}_{i (k l)j} +s_{13} \bar{R}_{i (j l) k} +s_{14} \bar{R}_{i (j k) l} \right] ~,
\end{align}
where the parentheses in the subscripts denote the symmetrization of indices, $A_{(ab)} \equiv (A_{ab} + A_{ba})/2$.
We have assumed all four external legs have incoming momentum $ p_{n} $ such that the conservation of momentum gives $ p_1 + p_2 + p_3 + p_4 =0 $, and the Mandelstam variable is defined as $ s_{nm} \equiv (p_n +p_m)^2 $ (see Fig.~\ref{fig:fourpoint}).
Specifically, in our case, we have
\begin{align}
    \left.  \mathcal{M}_{\phi^i \phi^i \to \phi^j \phi^j} \right|_\text{kin} = - R_N s + \frac{4}{3} R_N m_{\phi}^2(\bar{\Phi}_{\rm E}, \bar{\phi}^i) ~,
\end{align}
for $i\not = j$, where $s=s_{12}$ and
we have used the on-shell condition for $ \phi^i $. $ m_{\phi} $ is the effective mass of the Higgs which can be neglected.
By taking the metric limit $r = 1$, one can confirm that the result is consistent with the literature~\cite{Burgess:2009ea,Barbon:2009ya,Burgess:2010zq,Hertzberg:2010dc,Barvinsky:2009ii,Bezrukov:2010jz,Mikura:2021clt} albeit the contributions from spin-two graviton that we neglect.
In Ref.~\cite{Ema:2019fdd}, the frame independence of scattering amplitudes at vacuum is explicitly shown in the metric HI by comparing the results in the Jordan and the Einstein frames (see also Ref.~\cite{Hill:2020oaj}).
As we have already seen, this is automatically guaranteed in a geometric language because the Riemann tensor does not transform at vacuum.

Requiring the tree-level amplitude to be smaller than unity gives us the cutoff scale
\begin{align}\label{eq-cutoff-EC-tree}
    \Lambda_\text{E--C}^0 =|R_N|^{-1/2} = \frac{\sqrt{6}\Mpl}{\sqrt{1+12\xi \left(1+3r^2\xi \right)}} \sim
    \begin{cases}
        \dfrac{\Mpl}{r \xi} & \text{for} \quad \frac{1}{\sqrt{\xi}} \lesssim r \lesssim 1 ~, \\[1em]
        \dfrac{\Mpl}{\sqrt{\xi}} & \text{for} \quad 0 \leqslant r \lesssim \frac{1}{\sqrt{\xi}}~,
    \end{cases}
\end{align}
which is consistent with the known limits in the literature (see \textit{e.g.}, Ref.~\cite{Karananas:2022byw}).
In principle, we could also expand the field-space metric around some finite field value of Higgs, for instance, during inflation and get a field-value-dependent cutoff. We expect that we will get consistent result with that known in the literature~\cite{Bezrukov:2010jz,Mikura:2021clt,Karananas:2022byw}.

On the other hand, the potential depends on the field redefinition, and its cutoff scale is difficult to be determined unambiguously. Of course for $\lambda=0$ there is no relevant cutoff scale.
For non-zero $\lambda$, by performing the Weyl transformation on Eq.~\eqref{eq-Einstein-action-tilde1},
we recover the Planck mass and write down the action in the Einstein frame from
\begin{align}\label{eq-EC-tree-usual-Einstein}
    S= \int \dd^4 x \sqrt{-g_{\rm E}} &
    \Bigg[
        \frac{M_{\rm Pl}^2}{2} R_{\rm E}
        -\frac{1}{2} \frac{1}{\Omega_0^2}
        \left(
            \delta_{ij} + \frac{6\xi^2}{\Mpl^2} \frac{r^2}{\Omega_0^2} \phi_i \phi_j
        \right) g_{\rm E}^{\mu\nu} \partial_{\mu} \phi^i \partial_{\nu} \phi^j
        -\frac{V(\phi)}{\Omega_0^4}
    \Bigg] ~,
\end{align}
which is also good for convenient reading and comparison with Ref.~\cite{Shaposhnikov:2020gts}.
Here, $ R_{\rm E} $ is uniquely determined by $ g_{{\rm E}\mu\nu} $.
We may find the cutoff scale from the potential in this frame as, at most, $\sim \Mpl/\sqrt{\xi}$, which is always larger than that in Eq.~\eqref{eq-cutoff-EC-tree}.
Therefore, we determine Eq.~\eqref{eq-cutoff-EC-tree}
as the cutoff scale of the theory at the tree level in the vacuum.

Inflation takes place at the sufficiently large field
values of $\phi$, whose energy scale is
estimated as $ \left( \Lambda_\text{inf}^0 \right)^4 \simeq \left.V(\phi)/\Omega_0^4 \right|_{\phi \rightarrow \infty} = \lambda M_\text{Pl}^4/(4 \xi^2)$.
To obtain the observed amplitude of temperature fluctuations, there is a non-trivial relation between the non-minimal coupling $\xi$ and the Higgs four-point coupling $ \lambda $ as a function of $r$ as shown in Ref.~\cite{Shaposhnikov:2020gts}:
\begin{align}\label{eq-CMB-norm}
\frac{2 \lambda N_\text{e}^2}{\xi + 6 (r \xi)^2} \simeq 5.0 \times 10^{-7} \quad \Rightarrow  \quad
    \xi \simeq
    \begin{cases}
        4.5 \times 10^4 \dfrac{\sqrt{\lambda}}{r}
        & \text{for} \quad 2
        \times 10^{-6}~ \frac{1}{\sqrt{\lambda}}   \lesssim r \lesssim 1 ~, \\
        10^{10} \lambda & \text{for} \quad 0 \leqslant r \lesssim 2
        \times 10^{-6}~ \frac{1}{\sqrt{\lambda}} ~,
    \end{cases}
\end{align}
where we take the e-folding number at the CMB scale to be $N_\text{e}=55$.
Combining Eqs.~\eqref{eq-cutoff-EC-tree} and \eqref{eq-CMB-norm}, one finds that
unless $\lambda$ is extremely small, the cutoff scale is always much smaller than the Planck scale
and almost $ r $-independent
, \textit{i.e.,} $\Lambda_\text{E--C}^0 \sim 10^{-5} \Mpl/\sqrt{\lambda} $, which motivates us to seek the UV-extension.
In particular, in the metric limit, the unitarity problem arises during preheating~\cite{Ema:2016dny,Sfakianakis:2018lzf}.
The introduction of an $ R^2 $ term in the metric HI elegantly extends the cutoff to the Planck scale because the $ R^2 $ term induces a new scalar DoF, scalaron, that flattens the target space of the Higgs fields just as the linearization of a non-linear sigma model~\cite{Ema:2020zvg}.
The $ R^2 $ term can naturally arise from the quantum corrections to the action~\cite{Ema:2019fdd,Ema:2020zvg,Ema:2020evi} and it becomes important due to the renormalization group running with a large non-minimal coupling $ \xi $.
However, this is not the case in the Palatini formalism. The $ \bar{R}^2 $ term does not induces a new scalar DoF but turns the theory into a $ k $-essence theory~\cite{Enckell:2018hmo} whose cutoff is still lower than Planck scale.\footnote{
    Even if we take a different approach, \textit{i.e.}, introducing a new scalar DoF to flatten the target space, the higher-dimensional operators in the potential still ruins the effort to UV-extend the model~\cite{Mikura:2021clt}.
}
This can be seen from the coefficient of the $ X^2 $ term in Eq. (2.14) of Ref.~\cite{Enckell:2018hmo} that the cutoff around $ \phi =0 $ is $ \Lambda_{X^2} \sim \alpha^{-1/4} M_{\rm pl} $ $(\alpha\gg 1$).
This raises the question about how these two distinct results are smoothly connected in the context of E--C gravity since it is a general model containing both cases. If we consider the quantum corrections in E--C HI in the large-$ N $ limit, one expects that a corresponding ``$ R^2$  term'' should appear which controls whether a new scalar DoF appears and hopefully shows some hints on the way to UV-extend the Palatini HI model.\footnote{
    See Ref.~\cite{Mukohyama:2020lsu} and references therein for the attempts of UV-completion of $ k $-essence.
}

%%%%%%%%%%%%%%%%%%%%%%%%%%%%%%%%%%%%%%%%%%%%%%%%%%
\section{Quantum Corrections, New Degree of Freedom, and UV-Extension}
\label{Sec-3}
%%%%%%%%%%%%%%%%%%%%%%%%%%%%%%%%%%%%%%%%%%%%%%%%%%

In this section, we consider the quantum corrections to the action~\eqref{eq-total-action-jordan} at one-loop level in the large-$ N $ limit.
We will show that a new counter term is necessary to renormalize the theory.
Large-$ N $ analysis requires a resummation of a particular class of one-loop diagrams which are enhanced by $ N $, resulting in non-perturbative effect \textit{i.e.}, the emergence of a new pole in the scattering amplitude.
In some circumstances, the pole can be identified as a new dynamical degree of freedom, \textit{e.g.}, the scalaron~\cite{Ema:2019fdd}.
The coefficient of the counter term follows the renormalization group running, which determines the mass scale of the pole.
If the mass comes below the cutoff scale, the pole is naturally expected to become dynamical, which is the scalaron in our case, and possibly UV-extends the theory.
In the following, ``quantum correction'' is always referred to this meaning instead of those in the perturbative sense as the latter cannot extend the theory beyond the original cutoff. 
We only focus on the scalar sector in this system and will neglect the graviton (tensor) part when calculating the quantum corrections.
For this purpose, the action given in Eqs.~\eqref{eq-Jordan-conformal-mode-1} and \eqref{eq-def-A} is useful as the tensor part of the action is already separated in $\tilde g_{\mu\nu}$ and $\tilde R$.

\subsection{Quantum Corrections in Large $N$ \& $\xi$}

\begin{figure}[t]
	\centering
 	\includegraphics[width=0.7\linewidth]{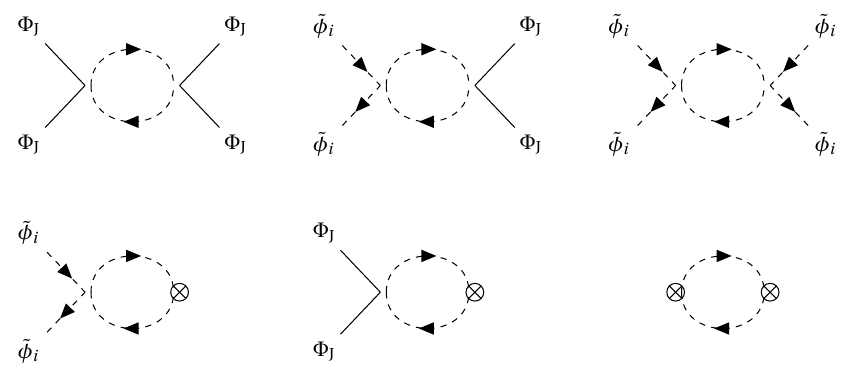}
	\caption{Divergent diagrams generated from Eq.~\eqref{eq-Jordan-conformal-mode-1} which contribute in the large-$N$ limit. The dashed line represents the  propagator of $\tilde \phi_i$  with the arrow being the flow of its flavor index $i$.
    All the diagrams contain a Higgs loop with the closed flavor flow, and thus they contribute at the leading order in the large $N$.
    The composite operator of $\tilde R + A_R + r A_\text{NY}$ is denoted by $\bigotimes$.
    The top three diagrams are renormalized from left to right by the cosmological constant $\Lambda/(9 \Mpl^4)$, the Higgs mass $m^2 / (3 \Mpl^2)$, and the Higgs four-point coupling $\lambda$.
    The bottom left and middle diagrams are renormalized by the non-minimal coupling $\xi$ and the wave-function renormalization of $\Phi_\text{J}$.
    However, the action \eqref{eq-Jordan-conformal-mode-1} does not have the corresponding term to renormalize the bottom right diagram.
    This is why we have to add the counter term \eqref{eq-counter-term-tilde} [or equivalently \eqref{eq-counter-term}].
}
	\label{fig:divergences}
\end{figure}

Throughout this paper, we are interested in the large-$ \xi $ case.
This is practically because the large $ \xi $ is typically required to obtain HI compatible with the current observation, and also because the coefficient of the terms like $ R_{\mu\nu} R^{\mu\nu} $ in the quantum corrections will become subdominant such that we can avoid the discussion of tensor modes or ghosts~\cite{Ema:2019fdd}.
We further assume the large-$ N $ limit where $ N $ is the number of real scalar fields in this system,
expecting $N=4$ for the SM Higgs is sufficiently ``large''.
The large-$N$ limit offers a perturbative expansion other than those based on the coupling constants, which allows us to investigate a UV completion of a nonlinear sigma model~\cite{Bardeen:1976zh,Brezin:1976ap,Dobado:1995qy,Dobado:1999xb,DAdda:1978vbw,DAdda:1978dle,Witten:1978bc,Arefeva:1980ms, Fujiwara:1984mp,Bando:1984ej,Bando:1987br,Nambu:1961tp,Nambu:1961fr,Gross:1974jv} including HI in E--C formalism [see Eq.~\eqref{eq-jordan-tilde2} and comments below].
The resummed one-loop effect in the large-$ N $ limit to the metric HI has already been studied in Ref.~\cite{Ema:2019fdd} where it is found that the leading order contribution results in a new scalar degree of freedom, the scalaron, which is nothing but the mixed Higgs-$ R^2 $ model~\cite{Ema:2017rqn}. In this paper, we follow the approach there and calculate the resummed one-loop effect to the E--C case.

In the large-$N$ limit, the dominant contributions stem from the Higgs loops where the flow of the Higgs-flavor index is closed.
In the dimensional regularization, the divergent diagrams
generated from \eqref{eq-redefine-Jordan} in the large-$N$ limit are shown in Fig.~\ref{fig:divergences}.
The divergences except for the bottom-right one are canceled out by the cosmological constant $\Lambda$, the mass $m$, the four-point coupling $\lambda$, the non-minimal coupling $\xi$, and the wave function renormalization of $\Phi_{\text{J}}$, respectively.
However, the last term lacks the corresponding term in \eqref{eq-Jordan-conformal-mode-1}, and hence the following counter term should be added to the action:
\begin{align}\label{eq-counter-term-tilde}
    \mathcal{L}_{\rm c.t.} =\alpha \qty( \tilde{R} + A_{R} + r A_\text{N--Y} )^2 ~,
\end{align}
or equivalently
\begin{align}\label{eq-counter-term}
    \mathcal{L}_\text{c.t.}
    = \alpha \qty[
        \bar R
        -
        \frac{r}{2} \nabla_\mu \qty( E^{\mu\nu\rho \sigma} T_{\nu\rho\sigma} )
    ]^2~,
\end{align}
in the Jordan frame.\footnote{
    Strictly speaking, there exists $q_{\alpha\beta\sigma}q^{\alpha\beta\sigma}$ in the parenthesis of Eq.~\eqref{eq-counter-term-tilde}.
    As mentioned earlier, we have dropped this term as it never contributes to the following discussion.
}
We have to restrict ourselves to $r > 1/\xi$ (or $r = 0$), \textit{i.e.}, $\xi_\eta > 1$ (or $\xi_\eta = 0$) for the consistency of our calculation.
This is because we only consider quantum corrections which are enhanced in the large-$\xi$ limit and drop others with order one couplings such as $\tilde \phi^2 \Box \Phi_{\text{J}} / \Phi_{\text{J}}$.
Note that we are mostly interested in the former branch $r > 1/\xi$ because the main purpose of this paper is to connect the metric and the Palatini HI smoothly by changing $r$ from unity to a smaller value continuously.
By extracting the divergent coefficient of the bare $\alpha$, one may readily obtain the running of $\alpha$ as
\begin{align}
    \label{eq-alpha-rge}
    \frac{\dd \alpha}{\dd \ln \mu} \simeq - \frac{N}{32 \pi^2} \xi^2 ~,
\end{align}
in the large $N$ and $\xi$.

The counter term \eqref{eq-counter-term-tilde} involves interaction terms of $\tilde T_\mu$ and $\tilde S_\mu$ whose power is higher than quadratic.
The Legendre transformation is useful to analyze such a system.
We introduce a new auxiliary field $ \gamma $ so that
\begin{align}
    S=
    & \int \dd^4 x
    \Bigg[
        \frac{\Phi_{\rm J}^2}{12} \tilde{R} +\frac{1}{2} \tilde{g}^{\mu\nu} \left( \partial_{\mu} \Phi_{\rm J} \partial_{\nu} \Phi_{\rm J} -\partial_{\mu} \tilde{\phi}^i \partial_{\nu} \tilde{\phi_i} \right)
        - \frac{\Phi_{\rm J}^2}{18} \left(
            \tilde T^2 - \frac{1}{16} \tilde{S}^2
        \right)
        - \left(
            \frac{\Phi_{\rm J}^2}{6\Mpl^2}
        \right)^2 V(\phi)
        -\frac{1}{2} \tilde{\phi}^2 \frac{\tilde{\Box}\Phi_{\rm J}}{\Phi_{\rm J}}
    \Bigg] \nonumber\\
    &+ \int \dd^4 x
    \left[
         \left(\tilde{R} + A_{R} + r A_\text{N--Y} - \gamma \right) L'(\gamma) +L(\gamma)
    \right] ~,
\end{align}
where
\begin{align}
    L(\gamma) \equiv \frac{\xi}{2} \tilde{\phi}^2 \gamma + \alpha \gamma^2 ~, \qquad
    L'(\gamma) = \frac{\dd L}{\dd \gamma} ~.
\end{align}
One can easily examine that $ \gamma $ coincides with $ \tilde{R} + A_{R} + r A_\text{N--Y} $ after solving the constraint equation for $\gamma$, which recovers the original action. This action can be further written in the following form which clearly shows the constraint for $ \tilde T_{\mu} $ and $ \tilde{S}_{\mu} $
\begin{align}\label{eq-Legendre-action-explicit-constraint}
    S= & \int \dd^4 x
    \Bigg[
        \frac{\Phi_{\rm J}^2}{12} \Omega^2 \tilde{R} +\frac{1}{2} \tilde{g}^{\mu\nu} \left( \partial_{\mu} \Phi_{\rm J} \partial_{\nu} \Phi_{\rm J} -\partial_{\mu} \tilde{\phi}^i \partial_{\nu} \tilde{\phi_i} \right) - \left( \frac{\Phi_{\rm J}^2}{6\Mpl^2} \right)^2 V(\phi) -\frac{1}{2} \tilde{\phi}^2 \frac{\tilde{\Box}\Phi_{\rm J}}{\Phi_{\rm J}}
        - \alpha \gamma^2
        -6 \qty( \frac{\xi}{2} \tilde \phi^2 + 2 \alpha \gamma ) \frac{\tilde{\Box}\Phi_{\rm J}}{\Phi_{\rm J}}
    \Bigg]  \nonumber\\
    &+ \int \dd^4 x
    \left[
        \left( 1-r^2 \right) \frac{\Phi_{\rm J}^2}{8} \Omega^2 \tilde{g}^{\mu\nu} \partial_{\mu} \ln \left| \Omega^2 \right| \partial_{\nu} \ln \left| \Omega^2 \right|
    \right] \nonumber \\
    &- \int \dd^4 x~ \frac{\Phi_{\rm J}^2}{18} \Omega^2 \left[
        \left( \tilde T_{\mu} + \frac{3}{2} \partial_{\mu} \ln \left| \Omega^2 \right| \right)^2
        - \frac{1}{16} \left( \tilde{S}_{\mu} - 6r \partial_{\mu} \ln \left| \Omega^2 \right| \right)^2
    \right] ~,
\end{align}
where
\begin{align}\label{eq-def-Omega}
    \Omega^2 \equiv 1+ 12\frac{L'(\gamma)}{\Phi_{\rm J}^2} =  1+ 12 \qty( \frac{\xi}{2} \frac{\tilde \phi^2}{\Phi_{\text{J}}^2} + 2 \alpha \frac{\gamma}{\Phi_{\text{J}}^2} )
    ~.
\end{align}
This action looks quite similar to Eq.~\eqref{eq-jordan-tilde2} except for the presence of a new field $\gamma$ in the interactions and the conformal factor.
Note that the target space is now spanned also by $\ln \left| \Omega^2 \right|$
in addition to the conformal mode and the Higgs fields,
since it is no longer uniquely determined by the latter fields.
Throughout this paper, we choose a ``healthy'' branch $\Omega^2 > 0$ so that the kinetic term of graviton is positive definite.
We also assume a positive $\alpha$ since otherwise the potential of the scalar sector would be unbounded below [see Eq.~\eqref{eq-Legendre-action-explicit-constraint} and also Eq.~\eqref{eq-Einstein-potential-sigma}].
Unless we set a fine tuned boundary condition at UV, the typical value of $ \alpha $ would be at least $ \alpha \sim \mathcal{O}(N\xi^2) \gg 1$ ~\cite{Ema:2020evi} because of the renormalization group running of $ \alpha $ given in \eqref{eq-alpha-rge} for a large non-minimal coupling $ \xi $ (which is the case in this paper)\footnote{
Note that, in the metric case ($ r=1 $)~\cite{He:2018mgb}, $ \alpha \sim \xi^2 $ corresponds to the Higgs-limit. If $ \alpha $ is treated as a free parameter and takes a larger value, the model approaches to the $ R^2 $-limit.
}.

The last line of Eq.~\eqref{eq-Legendre-action-explicit-constraint} shows the solutions to the constraints of $ \tilde T_{\mu} $ and $ \tilde{S}_{\mu} $, which can be dropped as long as the constraints are satisfied.
The first line is common for any value of $ r $ while the second line looks like a kinetic term for $ \Omega $ (or $ \gamma $) and it is essential for the discussion on how $ r $ affects the dynamics of the system.
We can go back to the tree-level HI by taking $ \alpha=0 $ in $ L(\gamma) $ as one can easily check.
For $ r=1 $ (or $ r=-1 $), \textit{i.e.,} the metric limit, the second line vanishes and
one can easily show that the first line coincides with the mixed Higgs-$ R^2 $ model, including an additional scalar DoF, \textit{i.e.,} scalaron, besides $ \phi^i $, by identifying $\gamma$ with $\sigma_J \Phi_{\text{J}} / (12 \alpha)$ in Eq.(4.16) of Ref.~\cite{Ema:2020zvg}.
In the exact limit of $ r=0 $, \textit{i.e.,} the Palatini limit,
as is known in the literature~\cite{Enckell:2018hmo}, the second line completely eliminates the new DoF and turns the model into a $k$-essence theory (see also Sec.~\ref{sec-new-dof}).
Yet, the main purpose of this paper is to connect the metric and Palatini HI by smoothly changing $r$ from $r = 1$ to a smaller value of $r$.
In this case, the result based on the action \eqref{eq-Legendre-action-explicit-constraint} has the restriction of $r > 1 / \xi$ as mentioned previously.
However, we do not need such an exact limit of $r = 0$ practically because the scalaron mass becomes Plankian for $r \to 1/\xi$ as we will see in the next section.

\subsection{Scalaron as New Degree of Freedom?}
\label{sec-new-dof}

The obtained action \eqref{eq-Legendre-action-explicit-constraint} involves a derivative acting on a new field $\gamma$ through the conformal factor \eqref{eq-def-Omega}, which may result in a new degree of freedom.
Let us start with a frame-independent criterion to see whether the action \eqref{eq-Legendre-action-explicit-constraint} for a general $r$ involves a new degree of freedom, \textit{i.e.,} scalaron, in addition to the conformal mode and the Higgs.

Owing to the frame independence of the geometrical quantities in which we are interested, one may choose a field basis where the calculation becomes particularly simple, which is again the Einstein frame.
In the same way as the previous section, the field transformation is given by
\begin{align} \label{eq-Einstein-metric}
    \Phi_{\rm E} \equiv \Omega \Phi_{\rm J} ~,
    \qquad
    g_{{\rm E}\mu\nu} = \frac{\Phi_{\rm E}^2}{6\Mpl^2} \tilde{g}_{\mu\nu} ~.
\end{align}
The resulting Einstein frame action becomes
\begin{align} \label{eq-Einstein-action-tilde2}
    S= \int \dd^4 x &
    \Bigg[
        \frac{\Phi_{\rm E}^2}{12} \tilde{R}
        +\frac{1}{2} \tilde{g}^{\mu\nu}
        \left(
            \partial_{\mu} \Phi_{\rm E} \partial_{\nu} \Phi_{\rm E} - \frac{\Phi_\text{E}^2/\Omega^2}{6\Mpl^2} \partial_{\mu} \phi_i \partial_{\nu} \phi^i
        \right)
        - \left( \frac{\Phi_\text{E}^2/\Omega^2}{6\Mpl^2} \right)^2 V(\phi) - \alpha \gamma^2\nonumber \\
    &- r^2 \frac{\Phi_\text{E}^2}{8} \tilde{g}^{\mu\nu} \partial_{\mu} \ln \left| \Omega^2 \right| \partial_{\nu} \ln \left| \Omega^2 \right|
    \Bigg] ~,
\end{align}
where we have already solved the constraint equations for $ \tilde T_{\mu} $ and $ \tilde{S}_{\mu} $.
The non-trivial kinetic term can be expressed by means of a matrix
    \begin{align}\label{eq-mixed-field-space-metric}
    \qty( G_{AB} ) = \begin{pmatrix}
        -1 & 0 & 0 \\
        0 &  \frac{\Phi_{\text{E}}^2 / \Omega^2}{6 \Mpl^2} \delta_{ij} & 0 \\
        0 & 0 &  r^2 \frac{\Phi_{\text{E}}^2}{4}
    \end{pmatrix}~,
\end{align}
as
$ -\sfrac{1}{2}\, G_{AB} \tilde{g}^{\mu\nu} \partial_{\mu} \varphi^A \partial_{\nu} \varphi^B
$ with $(\varphi^A) = (\Phi_\text{E}, \phi^i, \ln |\Omega^2|)$.
The determinant of $G_{AB}$ can be expressed as
\begin{align}
    \det \qty[ G_{AB} ]
    = - \qty( \frac{\Phi_{\text{E}}^2}{6 \Mpl^2} )^{N+1}
    \Omega^{-2N}
    \times \frac{3 \Mpl^2}{2} r^2~.
\end{align}
This result shows that the matrix $G_{AB}$ is non-degenerate for general $ r $, which means $ \Omega $ (or $ \gamma $) becomes a new scalar DoF, \textit{i.e.,} ``scalaron'', except for the exact limit of $ r=0 $ which, however, is not the main interest of the present analysis.
Note that the field redefinition never changes whether $G_{AB}$ is non-degenerate.
Therefore, we refer to the E--C HI with quantum corrections as the E--C mixed Higgs-scalaron inflation.

The fact that $ \ln |\Omega^2| $ becomes dynamical for general $r$ motivates us to define the scalaron as
\begin{align}
    \sigma_{\text{E}} = r \sqrt{\frac{3}{2}} \Mpl \ln \left| \Omega^2 \right|~,
\end{align}
where the subscript ``E'' implies that it is canonically normalized in the Einstein frame.
Together with the definition of $ \Omega $ in Eq.~\eqref{eq-def-Omega}, one can express $ \gamma $ in terms of $\sigma_{\text{E}}$ as follows
\begin{align}
    \gamma
    =
    \frac{\Phi_{\text{E}}^2 / \Omega^2}{6 \Mpl^2} \times \frac{\Mpl^2}{4 \alpha} \qty( \Omega^2 -  1 - \xi \frac{\phi^2}{\Mpl^2} )
    =
    \frac{\Phi_{\text{E}}^2}{6 \Mpl^2} \times e^{- \sqrt{\frac{2}{3}} \frac{\sigma_{\text{E}}}{r \Mpl} } \frac{\Mpl^2}{4 \alpha} \qty( e^{ \sqrt{\frac{2}{3}} \frac{\sigma_{\text{E}}}{r \Mpl} } -  1 - \xi \frac{\phi^2}{\Mpl^2} )~,
\end{align}
such that we can completely eliminate $\gamma$ in the action.
In terms of the scalaron $\sigma_{\text{E}}$, we can rewrite the action \eqref{eq-Einstein-action-tilde2} as
\begin{align} \label{eq-Einstein-action-tilde-sigma}
    S= \int \dd^4 x &
    \Bigg\{
        \frac{\Phi_{\rm E}^2}{12} \tilde{R}
        +\frac{1}{2} \tilde{g}^{\mu\nu}
        \left[
            \partial_{\mu} \Phi_{\rm E} \partial_{\nu} \Phi_{\rm E}
            - \frac{\Phi_\text{E}^2}{6\Mpl^2}
            \qty(
                e^{-  \sqrt{\frac{2}{3}} \frac{\sigma_{\text{E}}}{r \Mpl} }\partial_{\mu} \phi_i \partial_{\nu} \phi^i
                + \partial_\mu \sigma_{\text{E}} \partial_\nu \sigma_{\text{E}}
            )
        \right]
        - \left( \frac{\Phi_\text{E}^2}{6\Mpl^2} \right)^2 U(\phi, \sigma_{\text{E}})
    \Bigg\} ~,
\end{align}
where the potential is defined by
\begin{align} \label{eq-Einstein-potential-sigma}
    U (\phi, \sigma_{\text{E}})
    \equiv
    e^{-  2 \sqrt{\frac{2}{3}} \frac{\sigma_{\text{E}}}{r \Mpl} }
    \qty[
        V ( \phi )
        +
        \frac{\Mpl^4}{16 \alpha} \qty(
            e^{\sqrt{\frac{2}{3}} \frac{\sigma_{\text{E}}}{r \Mpl} } -  1 - \xi \frac{\phi^2}{\Mpl^2}
        )^2
    ]~.
\end{align}
The mass of ``scalaron'' at the vacuum, \textit{i.e.,} $\langle \phi_i \rangle = 0$ and $\langle \sigma_E \rangle = 0$, can be read off from the potential:
\begin{align} \label{eq-scalaron-mass-vac}
    \qty( \frac{\Phi_\text{E}^2}{6 \Mpl^2} )^2 U ( \phi, \sigma_\text{E} )
    \supset
    \qty( \frac{\Phi_\text{E}^2}{6} \frac{1}{12 \alpha r^2} ) \times \frac{1}{2} \qty( \frac{\Phi_\text{E}^2 \sigma_\text{E}^2}{6 \Mpl^2} )
    \quad
    \longrightarrow
    \quad
    m_{\sigma}^2 = \frac{\Mpl^2}{12 \alpha r^2}~,
\end{align}
where we assume that the vacuum energy, $\Lambda$, is negligible.
One may readily see this correspondence by absorbing the conformal mode $\Phi_\text{E}$ to $g_{\text{E}\mu\nu}$ and rewrite the action as follows:
\begin{align} \label{eq-Einstein-action-sigma}
    S = \int \dd^4 x \sqrt{ - g_\text{E}} \left[
        \frac{\Mpl^2}{2} R_\text{E}
        - \frac{1}{2} g_\text{E}^{\mu\nu} \left(
            \partial_\mu \sigma_\text{E} \partial_\nu \sigma_\text{E}
            +
            e^{ - \sqrt{\frac{2}{3}} \frac{\sigma_\text{E}}{r \Mpl} } \partial_\mu \phi_i \partial_\nu \phi^i
        \right)
        -
        U (\phi, \sigma_\text{E})
    \right]~.
\end{align}

\subsection{Cutoff Scale and Perturbativity}
\label{sec-cutoff}
So far we have confirmed that the new scalar DoF, \textit{i.e.,} scalaron, appears for general $r$ once we take into account quantum corrections to HI in the E--C formalism, and arrived at the E--C mixed Higgs-scalaron inflation given in Eqs.~\eqref{eq-Einstein-action-tilde-sigma} and \eqref{eq-Einstein-potential-sigma}.
The next question is to identify the cutoff scale (at vacuum) of the obtained action.\footnote{
    The cutoff scale during inflation can be different from that in vacuum.
    For instance, as discussed in Ref.~\cite{Bezrukov:2010jz}, the metric HI is consistent during inflation although the inflation scale is larger than the cutoff scale in vacuum  $\sim \Mpl / \xi $.
    However, during preheating after inflation, the problem arises again because of a large momentum of produced particles as mentioned in Introduction.
}
As shown in Eq.~\eqref{eq-cutoff-EC-tree}, the cutoff scale of HI without quantum corrections is  $\sim \Mpl/\xi $ for the metric formalism and $\sim \Mpl / \sqrt{\xi} $ for the Palatini formalism.
If the new cutoff scale with quantum corrections is larger than those without them, the obtained action can be regarded as a UV-extension.

According to the discussion in the previous section, the cutoff scale from the kinetic terms is characterized by the Riemann tensor of the target space.
Again, due to the frame covariance, we choose a field basis that can simplify the calculation, \textit{i.e.}, $(\varphi^A) \equiv (\Phi_{\rm E}, \phi^i, \sigma_{\rm E}) $ as in Eq.~\eqref{eq-Einstein-action-tilde-sigma}. As a result, the target space metric and the non-vanishing components of the Riemann tensor are as follows
\begin{align}
    \dd s^2 &= G_{AB} \dd \varphi^A \dd\varphi^B =  - \dd\Phi_{\rm E}^2 +\frac{\Phi_{\rm E}^2}{6\Mpl^2}
        \left(
            e^{ - \sqrt{\frac{2}{3}} \frac{\sigma_\text{E}}{r \Mpl} }\dd\phi^i \dd\phi_i + \dd\sigma_{\rm E}^2
        \right) ~, \label{eq-fieldspace-metric-mixed} \\
        R_{IKJL} &=
        - \frac{1 - r^2}{r^2 \Phi_\text{E}^2} \qty(
            G_{IJ} G_{KL} - G_{IL} G_{KJ}
        )~, \label{eq-general-riemann-mixed}
\end{align}
where $ A, B $ run from $ 0 $ (index for $ \Phi_{\rm E} $) to $ N+1 $ (index for $ \sigma_{\rm E} $) and $I, J, K, L$ run from $1$ to $N+1$.
At vacuum $ (\bar{\Phi}_{\rm E}, \bar{\phi}^i, \bar{\sigma}_{\rm E})= (\sqrt{6}\Mpl, \vec{0}, 0) $, the components of the Riemann tensor reduce to
\begin{align} \label{eq-riemann-mixed-vac}
    \bar{R}_{IKJL} = R_{N+s} \qty( \delta_{IJ} \delta_{KL} - \delta_{IL} \delta_{KJ} )~,
    \qquad R_{N+s} \equiv -\frac{1-r^2}{6r^2\Mpl^2} ~.
\end{align}
Again, the Riemann tensor at vacuum \eqref{eq-riemann-mixed-vac} is frame independent since the coordinate transformation at a vacuum is trivial, \textit{i.e.},
$\qty( \partial \Phi_\text{E} / \partial \Phi_\text{J} )_\text{vac} = 1$.

One can then follow the procedure in the HI case discussed previously, considering the four-point scattering $ \phi^i \phi^i \to \phi^j \phi^j \ (i \not= j) $ and $ \sigma \sigma \to \phi^i \phi^i $ (or $ \phi^i \phi^i \to \sigma \sigma $). The amplitudes of these processes can be calculated, from the interactions in the kinetic terms, as
\begin{align}
    \left. \mathcal{M}_{\phi^i \phi^i \to \phi^j \phi^j} \right|_\text{kin} &= -R_{N+s} s ~, \label{eq-iitojj-mixed} \\
    \left. \mathcal{M}_{\sigma \sigma \to \phi^i \phi^i} \right|_\text{kin} &= \frac{1}{3} R_{N+s} \left( -3 s +2 m_{\sigma}^2 \right) \label{eq-sigmasigmatojj} ~,
\end{align}
where we have again used the on-shell condition and taken $ m_{\phi}=0 $. In Eq.~\eqref{eq-iitojj-mixed}, we simply require this result is smaller than unity.
On the other hand, in Eq.~\eqref{eq-sigmasigmatojj}, the first term gives the same result as the previous case on the cutoff scale, while the second term gives an additional condition $ r\gtrsim (6\sqrt{3N}\xi)^{-1/2} $ for the model parameter by combining Eq.~\eqref{eq-scalaron-mass-vac} and taking $ \alpha \sim N\xi^2 $ such that the scalaron mass is within the cutoff of the theory.
As a result, one finds the frame-independent cutoff scale at vacuum from the non-canonical kinetic terms is
\begin{align}\label{eq-einstein-sigma-cutoff-1}
    \Lambda_{\text{E--C}} =|R_{N+s}|^{-1/2} = \frac{\sqrt{6}r}{\sqrt{|1-r^2|}} \Mpl ~.
\end{align}
As a special case, $ R_{N+s} $ vanishes for the metric limit $ r=1 $ (see discussion below for more details).
In the case of Palatini limit $ r \to 1/\xi $, $ R_{N+s} $ diverges (in proportional to $ \xi^2 $), leading to a low cutoff.
According to the general formula in Refs.~\cite{Nagai:2019tgi,Cohen:2021ucp}, one may also observe from Eqs.~\eqref{eq-fieldspace-metric-mixed} and \eqref{eq-general-riemann-mixed} that, for a positive $ \sigma_{\rm E} $ during inflation, the cutoff scale is the same as Eq.~\eqref{eq-einstein-sigma-cutoff-1}.

Besides, any operator with dimension higher than four in the potential also sets a scale above which strong coupling is expected.
As shown in the appendix~\ref{app-r-frame}, the potential~\eqref{eq-Einstein-potential-sigma} has a special feature so that one may always go to a particular field basis where the potential is exhausted up to four-point interactions.
This observation suggests that, in this model, the potential in a general field basis apparently features a cutoff scale, which, however, is originated from the field-space curvature.
Indeed, in the potential~\eqref{eq-Einstein-potential-sigma}, all the higher-dimensional operators come from the exponent involving $ \sigma_{\rm E} $, which implies a cutoff scale of $ \sim r\Mpl $, consistent with the field-space curvature.

It is instructive to take a closer look at the metric limit of $|r| = 1$.
In this case, we can take a field basis where the flatness of the field space is manifest.
By performing the following transformation
\begin{align}\label{eq-conformal-frame-transformation}
    \frac{\sigma_{\text{E}}}{\sqrt{6} \Mpl} - \ln \Phi_{\text{E}} = - \ln \qty( \Phi_\text{C} + \tilde \sigma )
    ~, \qquad
    \Phi_{\text{E}}^2 = \Phi_\text{C}^2 - \tilde \phi^2 - \tilde \sigma^2~,
\end{align}
one may rewrite the action in a simple form as follows:
\begin{align}\label{eq-r=1-conformal-frame}
    S = \int \dd^4 x &
    \Bigg[
        \frac{1}{12} \qty( \Phi_\text{C}^2 - \tilde \phi^2 - \tilde \sigma^2 ) \tilde{R}
        +\frac{1}{2} \tilde{g}^{\mu\nu}
        \left(
            \partial_{\mu} \Phi_{\rm C} \partial_{\nu} \Phi_{\rm C}
            - \partial_\mu \tilde \phi_i \partial_\nu \tilde \phi^i
            - \partial_\mu \tilde \sigma \partial_\nu \tilde \sigma
        \right)
        - U_\text{C} (\Phi_\text{C},\tilde \phi, \tilde \sigma)
    \Bigg] ~,
\end{align}
where the potential is defined by
\begin{align} \label{eq-potential-HR2-metric}
    U_\text{C} \qty( \Phi_\text{C}, \tilde \phi, \tilde \sigma )
    \equiv
    \frac{\lambda_\Lambda}{4} \left(\Phi_\text{C}+\tilde{\sigma}\right)^4
    +\frac{\lambda_m}{4} \left(\Phi_\text{C} +\tilde{\sigma} \right)^2 \tilde \phi^2
    + \frac{\lambda}{4} \tilde \phi^4
    + \frac{\lambda_\alpha}{4} \qty[
        \tilde \sigma \qty(\tilde \sigma + \Phi_\text{C}) + \frac{6 \xi + 1}{2} \tilde \phi^2
    ]^2 ~,
\end{align}
with
\begin{align}\label{eq-lambdas}
    \lambda_\Lambda \equiv \frac{\Lambda}{9 \Mpl^4}~, \qquad
    \lambda_m \equiv \frac{m^2}{3 \Mpl^2}~,
    \qquad
    \lambda_\alpha \equiv \frac{1}{36 \alpha}~.
\end{align}
This is nothing but the conformal frame of the mixed Higgs-$R^2$ model in the metric formalism~\cite{Ema:2020zvg}.
Somewhat surprisingly, the potential \eqref{eq-potential-HR2-metric} is exhausted up to four-point interactions in the basis where all the fields are canonically normalized and have no kinetic mixing, and moreover, no other interactions are required for renormalization even with quantum corrections from $\tilde \sigma$ and $\Phi_\text{C}$~\cite{Ema:2020evi}.\footnote{
    This remarkable feature is lost for general $r$.
}
Therefore, this theory is valid until the graviton comes into play, \textit{i.e.,} the cutoff scale $\sim M_\text{Pl}/\sqrt{N}$~\cite{Dvali:2007hz,Han:2004wt,Ema:2019fdd}.

To sum up, the cutoff scale of the E--C mixed Higgs-scalaron inflation for general $r$ can be estimated by the field-space curvature as follows:
\begin{align} \label{eq-cutoff-EC-quantum}
    \Lambda_\text{E--C}
    \sim
    \min \qty( \frac{\Mpl}{\sqrt{N}} , \frac{r\Mpl}{\sqrt{\qty|1 - r^2 |}} )~.
\end{align}
For $r \simeq 1$, the cutoff scale is $\Mpl / \sqrt{N}$ as it is dominated by the graviton, while the scalaron mass \eqref{eq-scalaron-mass-vac} is much smaller
$\Mpl / \sqrt{\alpha} \sim \Mpl / (\sqrt{N} \xi)$.
When we lower $r$ less than unity, \textit{i.e.,} $r \lesssim 1$, the cutoff scale decreases as $r \Mpl $, while the scalaron mass increases as $\Mpl / (r \sqrt{\alpha}) \sim \Mpl / (\sqrt{N} r \xi)$. At $r \sim \alpha^{-1/4} \sim N^{-1/4}\xi^{-1/2}$, the cutoff scale coincides with the scalaron mass, $\Lambda_\text{E--C} \sim m_{\sigma} \sim \Mpl / \alpha^{1/4} \sim \Mpl / (N^{1/4}\sqrt{\xi})$,
which is consistent with the discussion around Eq.~\eqref{eq-sigmasigmatojj}.
For $r \lesssim \alpha^{-1/4} $, the cutoff scale becomes smaller than the scalaron mass,
which poses a question on the reliability of the effective theory given in Eq.~\eqref{eq-Einstein-action-sigma} in this regime.
Meanwhile, when $ r \lesssim N^{1/4} \alpha^{-1/4} \sim 1/\sqrt{\xi} $ before the scalaron mass becomes too large, the new cutoff falls below that without quantum corrections, $ \Lambda_\text{E--C}^0 $.
Eventually, at $r \sim 1/\xi$, the scalaron mass reaches $ \sim \xi \Mpl / \sqrt{\alpha} \sim \Mpl/\sqrt{N} $, while the cutoff scale $\Mpl / \xi $ is smaller than $\Lambda_\text{Palatini} $.

Up to here, we have estimated the cutoff scale as \eqref{eq-cutoff-EC-quantum} above which the theory becomes strongly coupled and loses the perturbative unitarity.
However, this is not enough to ensure the perturbativity up to the cutoff scale since the potential~\eqref{eq-Einstein-potential-sigma} involves a large coupling $\xi$.
For $\alpha \sim \mathcal{O} (N\xi^2)$, the only term we need to take care is the four-point interaction of $\xi^2 \phi^4 / ( 16 \alpha )$.
As we are interested in the perturbative UV completion up to the cutoff scale, we require the running coupling at the cutoff scale to be $\alpha(\Lambda_\text{E--C}) \gtrsim \xi^2$.

Now we are ready to discuss whether the E--C HI is extended to UV once we take into account the counter term \eqref{eq-counter-term}, which is required by the renormalization of quantum corrections in the large-$N$ limit.
Comparing $\Lambda_\text{E--C}$ given in Eq.~\eqref{eq-cutoff-EC-quantum} to the result without quantum corrections $\Lambda_\text{E--C}^0$ in Eq.~\eqref{eq-cutoff-EC-tree}, one finds
\begin{align}
    \frac{\Lambda_\text{E--C}}{\Lambda_\text{E--C}^0} = \frac{\sqrt{1+12\xi \left(1+3r^2\xi\right)}}{\sqrt{\left|1-r^2\right|}} r \sim
    \begin{cases}
       \dfrac{6r^2\xi}{\sqrt{\left|1-r^2\right|}} > 1 & \quad \text{for} \quad \frac{1}{\sqrt{\xi}} \lesssim r \lesssim 1 ~, \\[1.em]
       \dfrac{\sqrt{12 \xi} r}{\sqrt{\left|1-r^2\right|}} < 1 & \quad \text{for} \quad \frac{1}{\xi} \lesssim r \lesssim \frac{1}{\sqrt{\xi}} ~.
    \end{cases}
\end{align}
For $ 1 /\sqrt{\xi} \lesssim r \lesssim 1 $,  the scalaron does UV-extend the original theory by participating in the dynamics, although the extension becomes smaller as the value of $ r $ decreases.
For $ 1/\xi \lesssim r \lesssim 1/\sqrt{\xi} $, however, the scalaron mass gets close to the cutoff scale and, at the same time, the cutoff of the new theory $ \Lambda_{\text{E--C}} $ falls below that of the old one $ \Lambda_\text{E--C}^0 $, which probably implies the breaking down of the current analysis and the correction from scalaron is not under control any more.
In the rest of this paper, we will not address this problem and focus on $ 1 /\sqrt{\xi} \lesssim r \lesssim 1 $.

\section{Mixed Higgs-Scalaron Inflation in Einstein--Cartan Gravity}\label{Sec-4}

\begin{figure}[t]
	\centering
 	\includegraphics[width=0.8\linewidth]{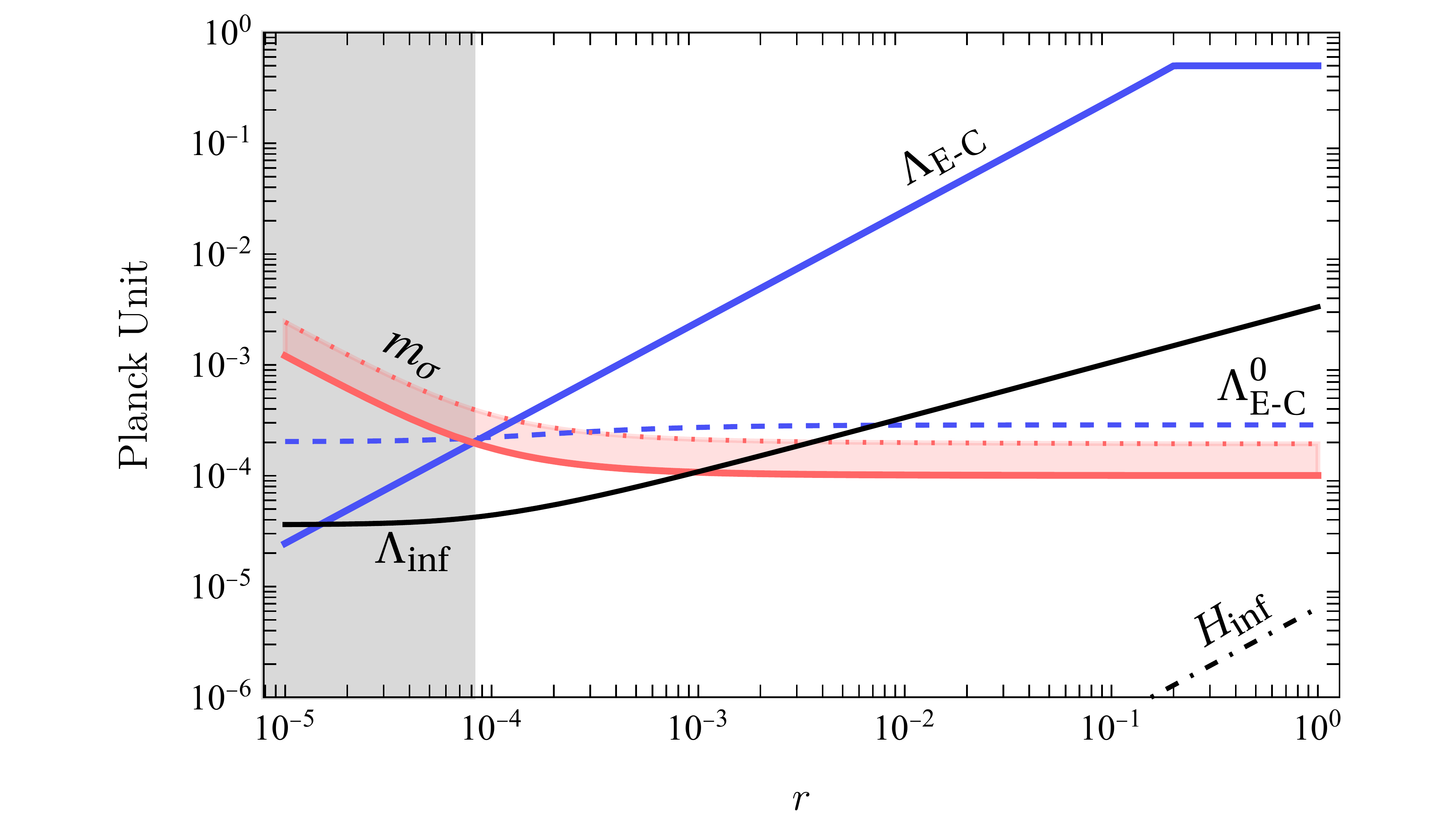}
	\caption{
        The $r$-dependence of various quantities are shown by imposing the CMB normalization in Eq.~\eqref{eq-CMB-norm}.
        Here we take the Higgs four-point coupling at the inflation scale to be $\lambda = 10^{-3}$ and the number of the scalar fields to be $N=4$.
        (\textbf{Black Solid}) the inflation scale $\Lambda_\text{inf}$ given in Eq.~\eqref{eq-infscale-cmb}.
        (\textbf{Black Dotdashed}) the Hubble parameter during inflation $H_\text{inf} $ given in Eq.~\eqref{eq-hubblescale-cmb}.
        (\textbf{Blue Dashed}) the cutoff scale of the E--C HI without quantum corrections, \textit{i.e.,} $\Lambda_\text{E--C}^0$ given in \eqref{eq-cutoff-EC-tree}.
        (\textbf{Blue Solid}) the cutoff scale with quantum corrections, \textit{i.e.,} $\Lambda_\text{E--C}$ given in Eq.~\eqref{eq-cutoff-EC-quantum}, which extends the validity of the E--C HI for $r \gtrsim 2 \times 10^{-5} / \sqrt{\lambda}$.
        (\textbf{Red Solid}) the scalaron mass in Eq.~\eqref{eq-scalaron-mass-vac} for $\alpha = N \xi^2$.
        (\textbf{Red Dotted})
        the maximum scalaron mass allowed by the perturbativity up to the cutoff scale $\Lambda_\text{E--C}$, \textit{i.e.}, $\alpha (\Lambda_\text{E--C}) = \xi^2$.
        (\textbf{Gray Shade})
        The scalaron mass exceeds the cutoff scale for $ r \lesssim 2\times 10^{-5}(N\lambda)^{-1/2} $ and hence the effective theory loses its control.
}
	\label{fig:comparison}
\end{figure}

In this section, we discuss the dynamics during inflation and the perspectives after inflation in the new model with quantum correction, namely the mixed Higgs-scalaron inflation in the Einstein--Cartan framework given in Eq.~\eqref{eq-Einstein-action-sigma}. As mentioned previously, the cutoff scale and the scalaron mass change as $ r $ varies.
The results of this section are summarized in Fig.~\ref{fig:comparison}.
In the following, we will show that the inflation scale also depends on $ r $.
Whether scalaron comes into play below the cutoff scale and the inflation scale significantly affects the dynamics of the theory. Besides, comparing the inflation scale and the cutoff is necessary to clarify the validity of the theory during inflation.
We will mainly discuss the regime $ 1/\sqrt{\xi} \lesssim r <1 $, because the metric limit $ r=1 $ has been well studied by many authors and the validity of the new theory is doubtable when $ 1/\xi \lesssim r \lesssim 1/\sqrt{\xi} $.
In the regime of our interest, the scalaron mass is smaller than the cutoff, so it UV-extends the theory as shown above, and it plays as important role as Higgs in the evolution of the system. Additionally, because of the known strong coupling problem during preheating in $ r =1 $ case in the absence of quantum correction~\cite{Ema:2016dny,Sfakianakis:2018lzf}, the UV cutoff is of the same significance after inflation.
Since the main purpose of this paper is not to study the reheating process in depth, we will just briefly comment on this aspect without investigating the possible high scales involved during (p)reheating that could jeopardize the validity of the theory.

\subsection{During Inflation}

\begin{figure}[t]
	\centering
    \includegraphics[width=0.52\linewidth]{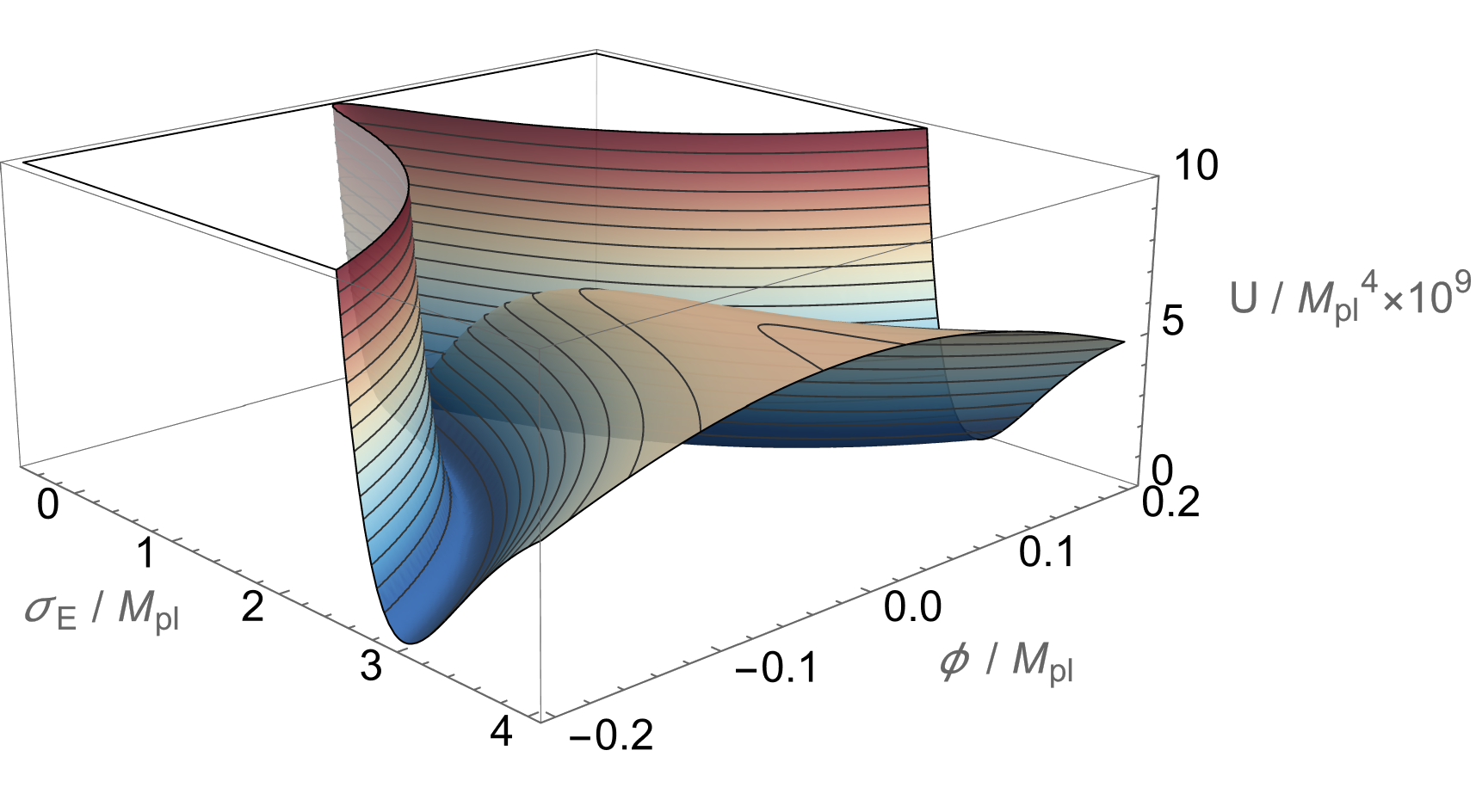}
 	\includegraphics[width=0.4\linewidth]{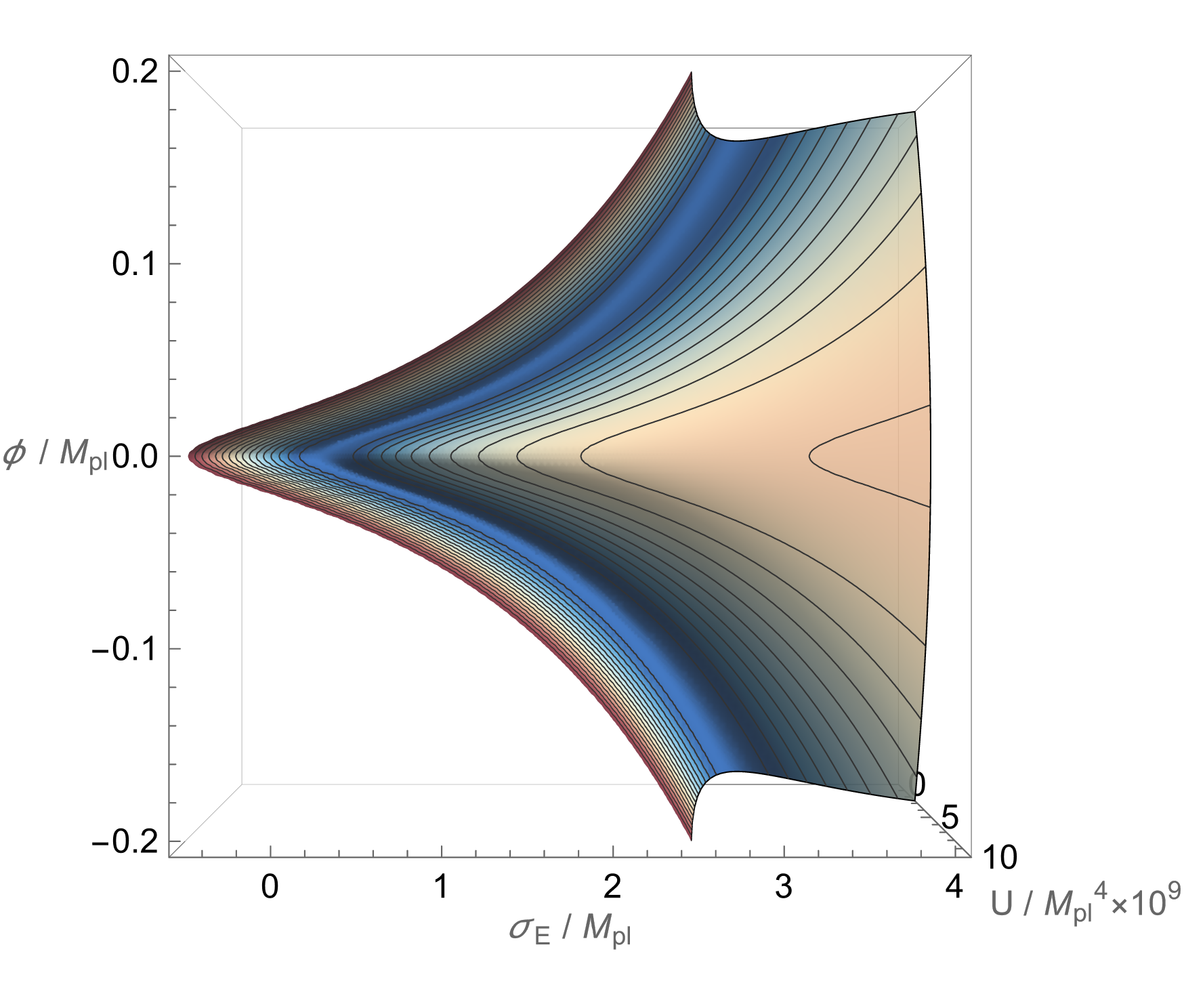}
	\caption{
        Potential~\eqref{eq-Einstein-potential-sigma} is shown with parameters $ \xi =3000 $, $ \alpha=\xi^2 $, $ \lambda = 10^{-3}$, and $ r =1/2 $. Left: side view. Right: top view.
        }
	\label{fig:potential}
\end{figure}

The inflation dynamics only involves the scalaron and the radial mode of the Higgs, so the situation is much simpler than that of (p)reheating where all the components of $ \phi^i $ are equally important. Inflation occurs in the ``valleys'' of the potentail~\eqref{eq-Einstein-potential-sigma}, which is similar to the metric case ($ r=1 $)~\cite{He:2018gyf}, as will be shown in the following. Fig.~\ref{fig:potential} shows a typical example of this potential from which this feature can be seen.
Inside the valley, the dynamics becomes essentially single field. More specifically, one can find out a light and a heavy direction in the valley and integrate out the latter to get an effective theory. See Appendix~\ref{app-heavy-light} for more detail.

The valleys can be well described by the local minima of $ \sigma_{\rm E} $ or $ \phi $ ($ \phi >0 $ as the radial direction, so one can focus on the positive $ \phi $ regime) during inflation
\begin{align}
    \frac{\partial U}{\partial \sigma_{\rm E}} &=0 \quad \rightarrow \quad e^{\sqrt{\frac{2}{3}}\frac{\sigma_{\rm E}}{r \Mpl}} - \xi \frac{\phi^2}{\Mpl^2} -1 = \frac{16\alpha V(\phi)/\Mpl^4}{1+ \xi \phi^2/\Mpl^2}  ~, \label{eq-sigma-minimum}\\
    \frac{\partial U}{\partial \phi} &=0 \quad \rightarrow \quad e^{\sqrt{\frac{2}{3}}\frac{\sigma_{\rm E}}{r \Mpl}} - \xi \frac{\phi^2}{\Mpl^2} -1 = \frac{4\alpha V'(\phi)}{\xi \Mpl^2 \phi}  ~.\label{eq-phi-minimum}
\end{align}
These two result coincide only when $ V(\phi) =0 $, which means that it is not a stationary point for the inflaton as the valleys are lifted up by the non-vanishing Higgs potential for the realization of slow-roll inflation. Fortunately, we can use either of them since they are approximately the same before the end of inflation, because the
following condition is satisfied
\begin{align}
    e^{\sqrt{\frac{2}{3}}\frac{\sigma_{\rm E}}{r \Mpl}} \sim \xi \frac{\phi^2}{\Mpl^2} ~\gtrsim ~\mathcal{O} (1) ~,
\end{align}
such that the potential~\eqref{eq-Einstein-potential-sigma} exhibits a plateau regime for the slow rolling of inflaton. One can show that, under this condition as well as even adopting the typical values from renormalization group running~\cite{Ema:2020evi}\footnote{In order to reconcile the present Universe as well as the SM Higgs, we need to assume that $\Lambda$ and $m$ are fine-tuned to be much smaller at the scalaron mass scale. But the discussion here does not change as long as they are not larger than the values predicted by the renormalization group running.}
\begin{align}
    \xi \gg 1 ~,~ \alpha \sim N \xi^2 ~,~ m \sim \frac{\Mpl}{\xi} ~,~ \Lambda \sim \frac{\Mpl^4}{\xi^4} ~,
\end{align}
the right hand sides of Eqs.~\eqref{eq-sigma-minimum} and \eqref{eq-phi-minimum} are negligible compared with the other parts
for a typical $ \lambda $ ($ \sim \mathcal{O}(10^{-3}) $ used in this paper) and
self-consistency. As a result, one can express, for instance, $ \sigma_{\rm E} $ in terms of $ \phi $, and the effective potential along the valley becomes
\begin{align}
    U_{\rm valley} \simeq e^{-2\sqrt{\frac{2}{3}}\frac{\sigma_{\rm E}(\phi)}{r\Mpl}} V(\phi) \simeq \frac{V(\phi)}{\left( 1+ \xi \frac{\phi^2}{\Mpl^2} \right)^2} ~,
\end{align}
which gives the
effective inflation scale and potential shape
of this model with the parameter choice considered in this paper, and they are the same as that in HI without quantum correction given in Eq.~\eqref{eq-EC-tree-usual-Einstein}.
Meanwhile, the kinetic terms
become
\begin{align}
    -\frac{1}{2} \left[ \frac{1}{1+ \xi \frac{\phi^2}{\Mpl^2}} + \frac{6 r^2 \xi^2 \phi^2/\Mpl^2}{\left( 1+ \xi \frac{\phi^2}{\Mpl^2} \right)^2} \right] \partial^\mu \phi \partial_\mu \phi ~,
\end{align}
which also coincides with that in Eq.~\eqref{eq-EC-tree-usual-Einstein}.
Therefore, the CMB normalization~\eqref{eq-CMB-norm} is still valid in this new model with quantum correction.
As a result, the inflation scale and Hubble parameter during inflation are
\begin{align}
    \Lambda_{\rm inf} &\simeq \Lambda_{\rm inf}^0 = \lambda^{1/4} \frac{ \Mpl }{\sqrt{2 \xi}}
    \sim
    \begin{cases}
        3 \times 10^{-3} \Mpl ~\sqrt{r}
        & \text{for} \quad  2 \times 10^{-6} \frac{1}{\sqrt{\lambda}} \lesssim r \lesssim 1 ~, \\
        4 \times 10^{-5} \Mpl~\qty( \dfrac{10^{-3}}{\lambda} )^{1/4} & \text{for} \quad 10^{-10}\frac{1}{\lambda} \leqslant r \lesssim 2 \times 10^{-6} \frac{1}{\sqrt{\lambda}} ~,
    \end{cases} \label{eq-infscale-cmb} \\
    H_{\rm inf} &\simeq H_{\rm inf}^0= \frac{\left( \Lambda_{\rm inf}^0 \right)^2}{\sqrt{3} \Mpl} \label{eq-hubblescale-cmb} ~,
\end{align}
although the lower case of Eq.~\eqref{eq-infscale-cmb} is irrelevant for the current discussion because we focus on $ r \gtrsim 1/\sqrt{\xi} $, \textit{i.e.}, $ r\gtrsim 2\times 10^{-5} \lambda^{-1/2} $ from Eq.~\eqref{eq-CMB-norm}.

We now compare $ \Lambda_{\rm inf} $ and $ \Lambda_\text{E--C} $ in the case where the scalaron mass is below the cutoff and the new cutoff $ \Lambda_{\text{E--C}} $ is above the one without quantum correction $ \Lambda_{\text{E--C}}^0 $, \textit{i.e.} $ 1/\sqrt{\xi} \lesssim r \lesssim 1 $ roughly corresponding to the upper case in Eq.~\eqref{eq-infscale-cmb},
which shows
\begin{align}\label{eq-compare-Lambda-inf-E-C}
    \frac{\Lambda_{\rm inf}}{\Lambda_\text{E--C}} \sim 3\times 10^{-3} ~ \sqrt{\frac{1}{r}-r} ~\lesssim ~ 0.1 \times \left( \frac{\lambda}{10^{-3}}\right)^{1/4} \quad \text{for} \quad \frac{1}{\sqrt{\xi}} \lesssim r \lesssim 1 ~,
\end{align}
where we have used Eq.~\eqref{eq-CMB-norm} to identify
the condition $ r \gtrsim 1/\sqrt{\xi}$
as $r \gtrsim 2\times 10^{-5} \lambda^{-1/2}$.
Therefore, inflation scale is at most the same as the cutoff at vacuum. The validity of inflation dynamics should be justified if the cutoff scale during inflation is larger than the inflation scale $ \Lambda_{\rm inf} $.
According to the discussion below Eq.~\eqref{eq-einstein-sigma-cutoff-1}, we expect that the cutoff scale around the non-trivial background inflationary dynamics is the same as $ \Lambda_\text{E--C} $, which means that the inflation scale is always smaller than the cutoff from the scattering of the fluctuations during mixed Higgs-scalaron inflation.
In addition, in the case $ r=1 $ with quantum correction, namely the mixed Higgs-$ R^2 $ model, it is also known that the theory is healthy up to Planck scale $ \Mpl/\sqrt{N} $ during (and after) inflation.

On the other hand, the scalaron mass can be larger than the Hubble scale during inflation as their ratio is given by
\begin{align}
    \frac{m_{\sigma}}{H_{\text{inf}}^0} \sim \dfrac{10}{\sqrt{N} r} \qty(\dfrac{\sqrt{N}\xi}{\sqrt{\alpha}})
    \left( \dfrac{\lambda}{10^{-3}}\right)^{-1/2} ~,
\end{align}
especially for a small $ r $ and a moderate $ N $. Therefore, one would like to integrate out the scalaron during inflation. In fact, the Higgs mass can also be larger than the Hubble scale, but one can always find a field basis where there exists one light mode as the inflaton and one heavy mode to be integrated out to successfully realize slow-roll inflation. This is not the main discussion here so we put more details in Appendix~\ref{app-heavy-light}.
Note that formally in $ r=0 $ case, integrating out the infinitely heavy scalaron should give a $ P(\phi, X) $ theory~\cite{Enckell:2018hmo,Antoniadis:2018ywb}.
One might wonder, in the current setup, whether the Palatini Higgs-$R^2$ inflation~\cite{Enckell:2018hmo,Antoniadis:2018ywb} is connected to the metric one by changing $ r $ continuously, since the theory tends to lose its control towards a smaller $ r \lesssim 1/\sqrt{\xi} $.
Indeed, the formal limit $ r=0 $ is lost in the current analysis due to the large-$ \xi $ limit and the validity of the resulting new theory, but as we have discussed,
the inflationary dynamics is expected to be described by an effective single-field theory without scalaron but with generally non-canonical kinetic terms.
Moreover, the inflaton trajectories (or the valleys) will become effectively parallel to the $ \phi $-axis (see Fig.~\ref{fig:potential}) as $ r $ gets smaller. In other words, the heavy direction is basically given by the scalaron direction. Therefore, we expect that, with $ r \ll 1 $, the resulting effective theory can well approximate the inflation dynamics of the formal $ r=0 $ limit. It would be fair to say that the Palatini Higgs-$R^2$ inflation is reproduced in the limit $ r \to 1/\sqrt{\xi} $.

\subsection{Perspectives of Post-Inflationary Dynamics}

As mentioned in Introduction, the motivation for UV extension of the metric HI is to avoid the strong coupling problem during preheating when the gauge bosons with momenta larger than the cutoff could be excited~\cite{Ema:2016dny,Sfakianakis:2018lzf}. Therefore, the UV extension during reheating is even more important for a healthy HI.

For a moment, let us neglect the possibility of the violent preheating known in the metric HI~\cite{Ema:2016dny,Sfakianakis:2018lzf}.
Suppose that the reheating is completed rather instantaneously after inflation, which is typically expected reheating after HI.
The would-be temperature of radiation in this case can be estimated by $T_\text{w.b.} \sim \Lambda_\text{inf}$.
If it exceeds the cutoff scale, the dynamics afterwards cannot be described by the same effective theory.
In the E-C HI without quantum corrections, one finds
\begin{align}\label{eq-temperature-condition}
    \frac{\Lambda_\text{inf}}{\Lambda_\text{E--C}^0} \sim
    10 \sqrt{r}\,  \qty( \frac{\lambda}{10^{-3}})^{1/2}
        > 1 \qquad \text{for} \quad r > 10^{-2} \qty(\frac{\lambda}{10^{-3}})^{-1}~.
\end{align}
This observation suggests that the E--C HI without quantum corrections would face the unitarity problem at the reheating stage for a moderately large $r$.
In the E--C mixed Higgs-scalaron inflation, this issue is resolved by the presence of scalaron, which can be seen from the comparison between $\Lambda_\text{inf}$ and $\Lambda_\text{E--C}$ in Eq.~\eqref{eq-compare-Lambda-inf-E-C}.
Hence, the unitarity problem after thermalization is solved in the E--C mixed Higgs-scalaron inflation.

On top of this, particles with much higher energy than $\Lambda_\text{inf}$ can be generated during preheating.
In the metric HI, it is known that the energy of produced particles can be as large as $\sqrt{\lambda} \Mpl \gg \Lambda_\text{inf}$. Nevertheless, the condition to avoid unitarity problem during violent preheating derived in Ref.~\cite{Ema:2021xhq} coincides with Eq.~\eqref{eq-temperature-condition}.
Again, the presence of the scalaron cures the preheating in the metric mixed Higgs-scalaron inflation~\cite{He:2018mgb}.
On the other hand, in the Palatini Higgs inflation, it is known that the violent preheating never occurs~\cite{Rubio:2019ypq}.
As these two theories are now connected by the change of $r$, we should be able to determine when the transition of these two pictures occurs.
Though we expect that the transition would occur around $r\sim 10^{-2} ( \lambda / 10^{-3})^{-1}$ based on Ref.~\cite{Ema:2021xhq},
the detailed investigation of particle production during (p)reheating in the new model is beyond the scope of this paper, so the discussion about that will be presented in a future work.

In summary, the one-loop correction to E--C HI can generally UV-extend the model to a higher scale for $ 1/\sqrt{\xi} \lesssim r \leq 1 $. Inflation dynamics is always healthy because the cutoff scale is larger than the inflation scale as shown. The cutoff scale at vacuum becomes larger than the tree-level model once the scalaron is taken into account.
However, whether (p)reheating process is also healthy after considering quantum corrections still needs investigation.

%%%%%%%%%%%%%%%%%%%%%%%%%%%%%%%%%%%%%%%%%%%%%%%%%%
\section{Conclusions and Discussion}
\label{Sec-5}
%%%%%%%%%%%%%%%%%%%%%%%%%%%%%%%%%%%%%%%%%%%%%%%%%%

In this paper, we have investigated the quantum corrections to HI within a general framework in E--C gravity, with the metric and the Palatini formalisms as special cases. In this framework, the two special cases mentioned above are realized by solving the constraint equations for the torsion tensor with corresponding parameters.
In each case, the phenomenology in the interested regimes have been intensively studied in the literature. The quantum corrections in the large-$ N $ limit results in an $ R^2 $ term in either case, which leads to a natural UV-extension to Planck scale in the metric limit known as the mixed Higgs-$ R^2 $ model but not in the Palatini case. This difference motivates us to investigate the quantum corrections to the general E--C case before solving any constraint in the system.

As a warmup, we started our discussion from the E--C HI without quantum corrections in Sec.~\ref{Sec-2}, and have demonstrated that the unitarity-violation scale is obtained from the curvature of the target-space extended to include the conformal mode of the metric, which is manifestly frame independent.
By changing the parameter $r$ from the Palatini limit $r = 0$ to the metric limit $|r| = 1$, we have confirmed that the target-space curvature at vacuum smoothly varies from
$ \sqrt{\xi}/\Mpl $ to $ \xi / \Mpl $.
Namely, $r$ can be regarded as a deformation parameter of the target space.
We have shown that the inflation energy scale consistent with the CMB normalization exceeds the cutoff scale for a sufficiently large $r$, implying the breakdown of the effective-field-theory description at the (p)reheating stage.

Then, we moved on to quantum corrections of the E--C HI in the large-$ N $ and $ \xi $ as shown in Sec.~\ref{Sec-3}.
This formal limit allows us to study quantum corrections from the scalar sector non-perturbatively while dropping those from the spin-two sector that appears much higher scale.
We have shown that a new scalar DoF, the scalaron $ \sigma $, is  generally induced by the quantum corrections whose mass is proportional to $ r^{-1} $.
Interestingly, the new target space including the scalaron is generally curved except for the metric limit $|r| = 1$, whose strong coupling scale is proportional to $r$.
For $r \lesssim 1 / \sqrt{\xi}$, the new cutoff scale $\Lambda_\text{E--C}$ falls below the cutoff without quantum corrections $\Lambda_\text{E--C}^0 $, and furthermore the scalaron mass exceeds the new cutoff, which casts doubts on the effective theory description.
Nevertheless, by analyzing the inflationary dynamics of the quantum corrected model, \textit{i.e.}, the mixed Higgs-scalaron inflation in E--C gravity, we can fairly say that the Palatini limit is reproduced as $ r \to 1/\sqrt{\xi} $ as we may integrate out the heavy mode which basically overlaps with the scalaron. This resulting effective theory is expected to be a good approximation for the formal Palatini limit of $ r=0 $.
For $ r\gtrsim 1/\sqrt{\xi} $, where our effective field theory is reliable, we always have $ m_{\sigma} \lesssim \Lambda_\text{E--C}^0 \lesssim \Lambda_\text{E--C} $, and hence the presence of the scalaron UV-extends the strong coupling scale up to $\Lambda_\text{E--C}$.
Moreover, we have shown that the new cutoff scale is always larger than the inflation scale for $ r\gtrsim 1/\sqrt{\xi} $.
This inequality admits the effective-field-theory description even at the reheating stage, at least after thermalization.

There are other attempts trying to UV-extend the Palatini limit where the large-$ N $ approach shows its limitation, for example, in Ref.~\cite{Mikura:2021clt} where the curved target space in Palatini Higgs inflation is embedded in a one-dimension-higher flat space to advoid the low cutoff from target space curvature.
However, the contribution from the higher-dimensional operators in the potential still shows low cutoff $ \sim \Mpl/\sqrt{\xi} $.
The UV-completion of the Palatini limit is still an open question worth inverstigating further. 

It is interesting to further investigate whether there is an attractor behavior of the non-minimal couplings that would always lead to a specific $ r $ such that a preferred model could be found in this general framework. In this paper, we restrict ourselves to the large-$N$ limit, and hence only regard $ r $ as a free parameter without running, but it could run in principle.
In that case, we expect there is a natural choice of $ r $ which can be considered as a preferred model or an attractor. In addition, a thorough investigation of the (p)reheating process is also desired for further understanding the role of the scalaron in E--C gravity, especially whether spikes and unitarity violation similar to the metric HI case~\cite{Ema:2016dny,He:2018mgb} generally exist and how the presence of the scalaron changes the situation.

%%%%%%%%%%%%%%%%%%%%%%%%%%%%%%%%%%%%%%%%%%%%%%%%%%
\section*{Acknowledgements}
%%%%%%%%%%%%%%%%%%%%%%%%%%%%%%%%%%%%%%%%%%%%%%%%%%
M.\,H.\, would like to thank Mikhail Shaposhnikov for the espresso and interesting discussion at CERN, and Tsutomu Kobayashi and Zhi-Bang Yao for useful discussion on modified gravity. 
We would like to thank Muzi Hong for useful comment.
K.\,K.\, was supported by JSPS KAKENHI Grant-in-Aid for Challenging Research (Exploratory) JP23K17687.
K.\,M.\, was supported by MEXT Leading Initiative for Excellent Young Researchers Grant No.\ JPMXS0320200430,
and by JSPS KAKENHI Grant No.\ 	JP22K14044.
All the Feynman diagrams in this paper have been produced by using \texttt{TikZ-FeynHand}~\cite{Dohse:2018vqo,Ellis:2016jkw}.

%%%%%%%%%%%%%%%%%%%%%%%%%%%%%%%%%%%%%%%%%%%%%%%%%%
\appendix
%%%%%%%%%%%%%%%%%%%%%%%%%%%%%%%%%%%%%%%%%%%%%%%%%%

%%%%%%%%%%
\section{``Planck Unit''}
\label{app-PhiE-vev}
%%%%%%%%%%

As implied by the Planck unit, only the ratio of dimensional quantities is physical, in particular, with respect to the ``Planck scale,'' where the graviton becomes strongly coupled.
Here we briefly show that the cutoff scale obtained from the target-space Riemann tensor is independent of the VEV of $\Phi_\text{E}$ in this sense.

Take the VEV of $\Phi_\text{E}$ to be $\sqrt{6} \Lambda_\text{G}$.
We expand the metric around the flat spacetime as
$\tilde g_{\mu\nu} = \qty[ e^h ]_{\mu\nu} = \eta_{\mu\nu} + h_{\mu\nu} + \cdots$ with $h_\mu^\mu = 0$ and $\partial^\mu h_{\mu\nu} = 0$.
The kinetic term of the graviton is obtained from
\begin{align}
    \int \dd^4 x\, \frac{\Phi_\text{E}^2}{12} \tilde R =
    \int \dd^4 x\, \frac{\Lambda_\text{G}^2}{8} \partial_\rho h_{\mu\nu} \partial^\rho h^{\mu\nu} + \cdots~.
\end{align}
After canonically normalizing $h_{\mu\nu}$, one finds that the cutoff scale for graviton interactions is $\Lambda_\text{G}$.
On the other hand, the normalization of Riemann tensor given in Eq.~\eqref{eq-riemann-HI} is given as
\begin{align}
    R_N = \frac{1 + 12 \xi ( 1 + 3 r^2 \xi)}{6 \Lambda_\text{G}^2}~.
\end{align}
This guarantees the following ratio does not depend on $\Lambda_\text{G}$ as expected:
\begin{align}
    \frac{\Lambda^0_\text{E--C}}{\Lambda_\text{G}}
    = \frac{\abs{R_\text{N}}^{-1/2}}{\Lambda_\text{G}}
    = \sqrt{ \frac{6}{1 + 12 \xi (1 + 3 r^2 \xi)} }~.
\end{align}
The same conclusion also holds for $\Lambda_\text{E--C}/ \Lambda_\text{G}$.

%%%%%%%%%%%%%%%%%%%%%%%%%%%%%%%%%%%%%%%%%%%%%%%%%%
\section{A Special Frame for General $ r $}\label{app-r-frame}
%%%%%%%%%%%%%%%%%%%%%%%%%%%%%%%%%%%%%%%%%%%%%%%%%%

In Sec.~\ref{sec-cutoff}, it has been shown that, for $ |r|=1 $, the action~\eqref{eq-Einstein-action-tilde-sigma} with potential~\eqref{eq-Einstein-potential-sigma} can be rewritten in the conformal frame~\eqref{eq-r=1-conformal-frame} where the kinetic terms for the scalar sector are all canonicalized and the potential is exhausted up to quartic interactions as Eq.~\eqref{eq-potential-HR2-metric}. Therefore, one can easily see that the cutoff scale of the theory is up to Planck scale. In this appendix, we are going to show that, for general $ r $, we can still find a frame where the potential share the same feature, but the kinetic terms for the scalar fields are generically not canonicalized. For later convenience, we denote this as the ``$ r $-frame".

In an analogous manner to Eq.~\eqref{eq-conformal-frame-transformation}, we define a Weyl transformation to the $ r $-frame which is characterized by the conformal mode $ \Phi_r $
\begin{align}
    \frac{\sigma_{\rm E}}{\sqrt{6}r M_{\rm Pl}} - \ln \Phi_{\rm E} = -\ln \left( \Phi_r +\tilde{\sigma} \right) ~, \qquad
    \Phi_{\text{E}}^2 = \Phi_r^2 - \tilde \phi^2 - \tilde \sigma^2~,
\end{align}
such that the action can be rewritten as
\begin{align}
    S = \int \dd^4 x &
    \Bigg[
        \frac{1}{12} \qty( \Phi_r^2 - \tilde \phi^2 - \tilde \sigma^2 ) \tilde{R}
        +\frac{1}{2} \tilde{g}^{\mu\nu} \left(
        K_{1\mu\nu} + r^2 K_{2\mu\nu} \right)
        - U_r (\Phi_r,\tilde \phi, \tilde \sigma)
    \Bigg] ~,
\end{align}
with the kinetic terms $ K_{1\mu\nu} $ and $ K_{2\mu\nu} $ given by
\begin{align}\label{eq-r-frame-target-space}
    K_{1\mu\nu} &\equiv -\partial_{\mu} \tilde{\phi}^i \partial_{\nu} \tilde{\phi}_i + \frac{\left( \tilde{\phi}^i \partial_{\mu}\tilde{\phi}_i + \tilde{\sigma}\partial_{\mu}\tilde{\sigma} - \Phi_r \partial_{\mu}\Phi_r \right)^2}{\Phi_r^2 -\tilde{\sigma}^2 - \tilde{\phi}^2} - \frac{\partial_{\mu}\left( \tilde{\sigma} +\Phi_r \right) \left[ \tilde{\phi}^2 \partial_{\nu} \left( \tilde{\sigma} +\Phi_r \right) -2\left( \tilde{\sigma} +\Phi_r \right) \tilde{\phi}_i \partial_{\nu} \tilde{\phi}^i \right]}{\left( \tilde{\sigma} +\Phi_r \right)^2} ~, \\
    K_{2\mu\nu} &\equiv -\frac{\left\{ \left( \tilde{\sigma} +\Phi_r \right) \tilde{\phi}_i \partial_{\nu} \tilde{\phi}^i +\left[ \left( \tilde{\sigma} +\Phi_r \right) \Phi_r -\tilde{\phi}^2 \right] \partial_{\mu} \tilde{\sigma} - \left[ \left( \tilde{\sigma} +\Phi_r \right) \tilde{\sigma} +\tilde{\phi^2} \right] \partial_{\mu} \Phi_r \right\}^2}{\left( \tilde{\sigma} +\Phi_r \right)^2 \left( \Phi_r^2 -\tilde{\sigma}^2 - \tilde{\phi}^2 \right)} ~,
\end{align}
and the potential
\begin{align}
    U_r \qty( \Phi_r, \tilde \phi, \tilde \sigma )
    \equiv
    \frac{\lambda_\Lambda}{4} \left(\Phi_r+\tilde{\sigma}\right)^4
    +\frac{\lambda_m}{4} \left(\Phi_r +\tilde{\sigma} \right)^2 \tilde \phi^2
    + \frac{\lambda}{4} \tilde \phi^4
    + \frac{\lambda_\alpha}{4} \qty[
        \tilde \sigma \qty(\tilde \sigma + \Phi_r) + \frac{6 \xi + 1}{2} \tilde \phi^2
    ]^2 ~,
\end{align}
where the couplings are the same as Eq.~\eqref{eq-lambdas}. As can be easily checked, the kinetic terms become canonical by taking $ |r|=1 $. Therefore, the curvature of the target space is generally non-vanishing even though the potential contains interactions only up to quartic order. As long as the couplings in this potential is small enough, there is no violation of unitarity from the potential.  Since the curvature of the target space is frame-independent, this means that the cutoff scale of the theory in the $ r $-frame is determined by the curvature of target space calculated in the Einstein frame~\eqref{eq-einstein-sigma-cutoff-1}, which shows the self-consistency of our analysis. Even if we remove all the higher-dimensional operators in the potential by using the $ r $-frame, the only cutoff scale coming from the target space should lead to the same cutoff as in other frames.

%%%%%%%%%%%%%%%%%%%%%%%%%%%%%%%%%%%%%%%%%%%%%%%%%%
\section{Heavy and Light Modes during Inflation}\label{app-heavy-light}
%%%%%%%%%%%%%%%%%%%%%%%%%%%%%%%%%%%%%%%%%%%%%%%%%%

The purpose of this appendix is to show that there is a heavy and a light mode during inflation in this new model~\eqref{eq-Einstein-action-sigma}. During inflation, only the radial direction of $ \phi^i $ is relevant, \textit{i.e.} $ \phi >0 $, so we will neglect the angular part in the following discussion for simplicity. Similar to the metric case ($ r=1 $), there are two valleys in the two-field potential~\eqref{eq-Einstein-potential-sigma} where inflation actually occurs. Fig.~\ref{fig:potential} shows a typical example of this potential. The location of these two valleys are well approximated by the local minima of $ \sigma_{\rm E} $ or $ \phi $ during inflation, as shown in Eqs.~\eqref{eq-sigma-minimum} and \eqref{eq-phi-minimum}, with the potential $ V(\phi) $ negligible in the interested parameter space.
To simplify the calculation, we consider the regime well before the end of inflation, \textit{i.e.}, $ \exp (\sqrt{\frac{2}{3}}\frac{\sigma}{r \Mpl}) \gg 1 $, which is sufficient for our purpose here. For a given position $ (\sigma_0, \phi_0) $ in the valley,
\begin{align}
    e^{\sqrt{\frac{2}{3}}\frac{\sigma_0}{r \Mpl}} \simeq \xi \frac{\phi_0^2}{\Mpl^2} \gg 1 ~,
\end{align}
one can calculate the mass of scalaron and Higgs which are given as
\begin{align}\label{eq-masses-valley}
    m_{\sigma}^2 \simeq \frac{\Mpl^2}{12\alpha r^2} ~, ~
    m_{\phi}^2 \simeq \frac{\Mpl^4}{2\alpha \phi_0^2} ~.
\end{align}
The effective mass of Higgs is $ \phi $-dependent. During inflation, $ \phi \gg \Mpl/\sqrt{\xi} $, so we find that
\begin{align}
    m_{\phi} \ll \frac{\Mpl}{\sqrt{\xi}} ~.
\end{align}
As inflation proceeds, $ \phi $ decreases such that $ m_{\phi} $ increases. Once inflation is about to end, the approximation to obtain Eq.~\eqref{eq-masses-valley} breaks down so $ m_{\phi} $ will not simply diverge as $ \phi \to 0 $.

The scalaron mass and the Higgs mass can both be much larger than the Hubble scale during inflation, but the valleys show that there exists a light direction that support slow-roll inflation while the heavy direction can be integrated out in principle since its mass is much larger than the interested scale. It is useful to expand the potential $ (\sigma_0 + \delta \sigma, \phi_0+ \delta \phi) $ up to second order and neglect $ V(\phi) $, such that
\begin{align}
    U \simeq \frac{\Mpl^2}{24\alpha r^2} \left( \delta \sigma - \frac{\sqrt{6}r \Mpl}{\phi_0} \delta \phi \right)^2 ~.
\end{align}
The expression shows the heavy direction and the light direction is absent because it is massless due to the approximation $ V(\phi)=0 $. In the presence of small $ V(\phi) $, the light mode obtains a small mass. As a result, one can always rotate the fields $\sigma_{\rm E}$ and $ \phi $ in a field-value-dependent way to obtain a light (inflation direction) and a heavy modes, although both Higgs and scalaron can be heavier than the Hubble scale during inflation.

%%%%%%%%%%%%%%%%%%%%%%%%%%%%%%%%%%%%%%%%%%%%%%%%%%
\small
\bibliographystyle{utphys}
\bibliography{ref}

\providecommand{\href}[2]{#2}\begingroup\raggedright\begin{thebibliography}{100}

\bibitem{Starobinsky:1980te}
A.~A. Starobinsky, ``{A New Type of Isotropic Cosmological Models Without Singularity},'' \href{http://dx.doi.org/10.1016/0370-2693(80)90670-X}{{\em Phys. Lett. B} {\bfseries 91} (1980) 99--102}.

\bibitem{Guth:1980zm}
A.~H. Guth, ``{The Inflationary Universe: A Possible Solution to the Horizon and Flatness Problems},'' \href{http://dx.doi.org/10.1103/PhysRevD.23.347}{{\em Phys. Rev. D} {\bfseries 23} (1981) 347--356}.

\bibitem{Sato:1980yn}
K.~Sato, ``{First Order Phase Transition of a Vacuum and Expansion of the Universe},'' {\em Mon. Not. Roy. Astron. Soc.} {\bfseries 195} (1981) 467--479.

\bibitem{Mukhanov:1981xt}
V.~F. Mukhanov and G.~V. Chibisov, ``{Quantum Fluctuations and a Nonsingular Universe},'' {\em JETP Lett.} {\bfseries 33} (1981) 532--535.

\bibitem{Linde:1981mu}
A.~D. Linde, ``{A New Inflationary Universe Scenario: A Possible Solution of the Horizon, Flatness, Homogeneity, Isotropy and Primordial Monopole Problems},'' \href{http://dx.doi.org/10.1016/0370-2693(82)91219-9}{{\em Phys. Lett. B} {\bfseries 108} (1982) 389--393}.

\bibitem{Albrecht:1982wi}
A.~Albrecht and P.~J. Steinhardt, ``{Cosmology for Grand Unified Theories with Radiatively Induced Symmetry Breaking},'' \href{http://dx.doi.org/10.1103/PhysRevLett.48.1220}{{\em Phys. Rev. Lett.} {\bfseries 48} (1982) 1220--1223}.

\bibitem{Sato:2015dga}
K.~Sato and J.~Yokoyama, ``{Inflationary cosmology: First 30+ years},'' \href{http://dx.doi.org/10.1142/S0218271815300256}{{\em Int. J. Mod. Phys. D} {\bfseries 24} no.~11, (2015) 1530025}.

\bibitem{BENNETT1993409}
C.~Bennett, N.~Boggess, E.~Cheng, M.~Hauser, T.~Kelsall, J.~Mather, S.~Moseley, T.~Murdock, R.~Shafer, R.~Silverberg, G.~Smooth, R.~Weiss, and E.~Wright, ``{Scientific results from COBE},'' \href{http://dx.doi.org/https://doi.org/10.1016/0273-1177(93)90150-A}{{\em Advances in Space Research} {\bfseries 13} no.~12, (1993) 409--423}.

\bibitem{Komatsu_2011}
E.~Komatsu, K.~M. Smith, J.~Dunkley, C.~L. Bennett, B.~Gold, G.~Hinshaw, N.~Jarosik, D.~Larson, M.~R. Nolta, L.~Page, D.~N. Spergel, M.~Halpern, R.~S. Hill, A.~Kogut, M.~Limon, S.~S. Meyer, N.~Odegard, G.~S. Tucker, J.~L. Weiland, E.~Wollack, and E.~L. Wright, ``{Seven-Year Wilkinson Microwave Anisotropy Probe (WMAP) Observations: Cosmological Interpretation},'' \href{http://dx.doi.org/10.1088/0067-0049/192/2/18}{{\em The Astrophysical Journal Supplement Series} {\bfseries 192} no.~2, (Jan, 2011) 18}.

\bibitem{Planck:2018jri}
{\bfseries Planck} Collaboration, Y.~Akrami {\em et~al.}, ``{Planck 2018 results. X. Constraints on inflation},'' \href{http://dx.doi.org/10.1051/0004-6361/201833887}{{\em Astron. Astrophys.} {\bfseries 641} (2020) A10}, \href{http://arxiv.org/abs/1807.06211}{{\ttfamily arXiv:1807.06211 [astro-ph.CO]}}.

\bibitem{Cervantes-Cota:1995ehs}
J.~L. Cervantes-Cota and H.~Dehnen, ``{Induced gravity inflation in the standard model of particle physics},'' \href{http://dx.doi.org/10.1016/0550-3213(95)00128-X}{{\em Nucl. Phys. B} {\bfseries 442} (1995) 391--412}, \href{http://arxiv.org/abs/astro-ph/9505069}{{\ttfamily arXiv:astro-ph/9505069}}.

\bibitem{Bezrukov:2007ep}
F.~L. Bezrukov and M.~Shaposhnikov, ``{The Standard Model Higgs boson as the inflaton},'' \href{http://dx.doi.org/10.1016/j.physletb.2007.11.072}{{\em Phys. Lett. B} {\bfseries 659} (2008) 703--706}, \href{http://arxiv.org/abs/0710.3755}{{\ttfamily arXiv:0710.3755 [hep-th]}}.

\bibitem{Horn:2020wif}
B.~Horn, ``{The Higgs field and early universe cosmology: a (brief) review},'' \href{http://dx.doi.org/10.3390/physics2030028}{{\em MDPI Physics} {\bfseries 2} no.~3, (2020) 503--520}, \href{http://arxiv.org/abs/2007.10377}{{\ttfamily arXiv:2007.10377 [hep-ph]}}.

\bibitem{Ema:2016dny}
Y.~Ema, R.~Jinno, K.~Mukaida, and K.~Nakayama, ``{Violent Preheating in Inflation with Nonminimal Coupling},'' \href{http://dx.doi.org/10.1088/1475-7516/2017/02/045}{{\em JCAP} {\bfseries 02} (2017) 045}, \href{http://arxiv.org/abs/1609.05209}{{\ttfamily arXiv:1609.05209 [hep-ph]}}.

\bibitem{Sfakianakis:2018lzf}
E.~I. Sfakianakis and J.~van~de Vis, ``{Preheating after Higgs Inflation: Self-Resonance and Gauge boson production},'' \href{http://dx.doi.org/10.1103/PhysRevD.99.083519}{{\em Phys. Rev. D} {\bfseries 99} no.~8, (2019) 083519}, \href{http://arxiv.org/abs/1810.01304}{{\ttfamily arXiv:1810.01304 [hep-ph]}}.

\bibitem{Burgess:2009ea}
C.~P. Burgess, H.~M. Lee, and M.~Trott, ``{Power-counting and the Validity of the Classical Approximation During Inflation},'' \href{http://dx.doi.org/10.1088/1126-6708/2009/09/103}{{\em JHEP} {\bfseries 09} (2009) 103}, \href{http://arxiv.org/abs/0902.4465}{{\ttfamily arXiv:0902.4465 [hep-ph]}}.

\bibitem{Barbon:2009ya}
J.~L.~F. Barbon and J.~R. Espinosa, ``{On the Naturalness of Higgs Inflation},'' \href{http://dx.doi.org/10.1103/PhysRevD.79.081302}{{\em Phys. Rev. D} {\bfseries 79} (2009) 081302}, \href{http://arxiv.org/abs/0903.0355}{{\ttfamily arXiv:0903.0355 [hep-ph]}}.

\bibitem{Burgess:2010zq}
C.~P. Burgess, H.~M. Lee, and M.~Trott, ``{Comment on Higgs Inflation and Naturalness},'' \href{http://dx.doi.org/10.1007/JHEP07(2010)007}{{\em JHEP} {\bfseries 07} (2010) 007}, \href{http://arxiv.org/abs/1002.2730}{{\ttfamily arXiv:1002.2730 [hep-ph]}}.

\bibitem{Hertzberg:2010dc}
M.~P. Hertzberg, ``{On Inflation with Non-minimal Coupling},'' \href{http://dx.doi.org/10.1007/JHEP11(2010)023}{{\em JHEP} {\bfseries 11} (2010) 023}, \href{http://arxiv.org/abs/1002.2995}{{\ttfamily arXiv:1002.2995 [hep-ph]}}.

\bibitem{Barvinsky:2009ii}
A.~O. Barvinsky, A.~Y. Kamenshchik, C.~Kiefer, A.~A. Starobinsky, and C.~F. Steinwachs, ``{Higgs boson, renormalization group, and naturalness in cosmology},'' \href{http://dx.doi.org/10.1140/epjc/s10052-012-2219-3}{{\em Eur. Phys. J. C} {\bfseries 72} (2012) 2219}, \href{http://arxiv.org/abs/0910.1041}{{\ttfamily arXiv:0910.1041 [hep-ph]}}.

\bibitem{Bezrukov:2010jz}
F.~Bezrukov, A.~Magnin, M.~Shaposhnikov, and S.~Sibiryakov, ``{Higgs inflation: consistency and generalisations},'' \href{http://dx.doi.org/10.1007/JHEP01(2011)016}{{\em JHEP} {\bfseries 01} (2011) 016}, \href{http://arxiv.org/abs/1008.5157}{{\ttfamily arXiv:1008.5157 [hep-ph]}}.

\bibitem{Giudice:2010ka}
G.~F. Giudice and H.~M. Lee, ``{Unitarizing Higgs Inflation},'' \href{http://dx.doi.org/10.1016/j.physletb.2010.10.035}{{\em Phys. Lett. B} {\bfseries 694} (2011) 294--300}, \href{http://arxiv.org/abs/1010.1417}{{\ttfamily arXiv:1010.1417 [hep-ph]}}.

\bibitem{Barbon:2015fla}
J.~L.~F. Barbon, J.~A. Casas, J.~Elias-Miro, and J.~R. Espinosa, ``{Higgs Inflation as a Mirage},'' \href{http://dx.doi.org/10.1007/JHEP09(2015)027}{{\em JHEP} {\bfseries 09} (2015) 027}, \href{http://arxiv.org/abs/1501.02231}{{\ttfamily arXiv:1501.02231 [hep-ph]}}.

\bibitem{Ema:2017rqn}
Y.~Ema, ``{Higgs Scalaron Mixed Inflation},'' \href{http://dx.doi.org/10.1016/j.physletb.2017.04.060}{{\em Phys. Lett. B} {\bfseries 770} (2017) 403--411}, \href{http://arxiv.org/abs/1701.07665}{{\ttfamily arXiv:1701.07665 [hep-ph]}}.

\bibitem{Lee:2018esk}
H.~M. Lee, ``{Light inflaton completing Higgs inflation},'' \href{http://dx.doi.org/10.1103/PhysRevD.98.015020}{{\em Phys. Rev. D} {\bfseries 98} no.~1, (2018) 015020}, \href{http://arxiv.org/abs/1802.06174}{{\ttfamily arXiv:1802.06174 [hep-ph]}}.

\bibitem{Koshelev:2020fok}
A.~S. Koshelev and A.~Tokareva, ``{Non-local self-healing of Higgs inflation},'' \href{http://dx.doi.org/10.1103/PhysRevD.102.123518}{{\em Phys. Rev. D} {\bfseries 102} (2020) 123518}, \href{http://arxiv.org/abs/2006.06641}{{\ttfamily arXiv:2006.06641 [hep-th]}}.

\bibitem{Salvio:2015kka}
A.~Salvio and A.~Mazumdar, ``{Classical and Quantum Initial Conditions for Higgs Inflation},'' \href{http://dx.doi.org/10.1016/j.physletb.2015.09.020}{{\em Phys. Lett. B} {\bfseries 750} (2015) 194--200}, \href{http://arxiv.org/abs/1506.07520}{{\ttfamily arXiv:1506.07520 [hep-ph]}}.

\bibitem{Netto:2015cba}
T.~d.~P. Netto, A.~M. Pelinson, I.~L. Shapiro, and A.~A. Starobinsky, ``{From stable to unstable anomaly-induced inflation},'' \href{http://dx.doi.org/10.1140/epjc/s10052-016-4390-4}{{\em Eur. Phys. J. C} {\bfseries 76} no.~10, (2016) 544}, \href{http://arxiv.org/abs/1509.08882}{{\ttfamily arXiv:1509.08882 [hep-th]}}.

\bibitem{Liu:2018hno}
L.-H. Liu, T.~Prokopec, and A.~A. Starobinsky, ``{Inflation in an effective gravitational model and asymptotic safety},'' \href{http://dx.doi.org/10.1103/PhysRevD.98.043505}{{\em Phys. Rev. D} {\bfseries 98} no.~4, (2018) 043505}, \href{http://arxiv.org/abs/1806.05407}{{\ttfamily arXiv:1806.05407 [gr-qc]}}.

\bibitem{Calmet:2016fsr}
X.~Calmet and I.~Kuntz, ``{Higgs Starobinsky Inflation},'' \href{http://dx.doi.org/10.1140/epjc/s10052-016-4136-3}{{\em Eur. Phys. J. C} {\bfseries 76} no.~5, (2016) 289}, \href{http://arxiv.org/abs/1605.02236}{{\ttfamily arXiv:1605.02236 [hep-th]}}.

\bibitem{Ghilencea:2018rqg}
D.~M. Ghilencea, ``{Two-loop corrections to Starobinsky-Higgs inflation},'' \href{http://dx.doi.org/10.1103/PhysRevD.98.103524}{{\em Phys. Rev. D} {\bfseries 98} no.~10, (2018) 103524}, \href{http://arxiv.org/abs/1807.06900}{{\ttfamily arXiv:1807.06900 [hep-ph]}}.

\bibitem{Ema:2019fdd}
Y.~Ema, ``{Dynamical Emergence of Scalaron in Higgs Inflation},'' \href{http://dx.doi.org/10.1088/1475-7516/2019/09/027}{{\em JCAP} {\bfseries 09} (2019) 027}, \href{http://arxiv.org/abs/1907.00993}{{\ttfamily arXiv:1907.00993 [hep-ph]}}.

\bibitem{Ema:2020zvg}
Y.~Ema, K.~Mukaida, and J.~van~de Vis, ``{Higgs inflation as nonlinear sigma model and scalaron as its $\sigma$-meson},'' \href{http://dx.doi.org/10.1007/JHEP11(2020)011}{{\em JHEP} {\bfseries 11} (2020) 011}, \href{http://arxiv.org/abs/2002.11739}{{\ttfamily arXiv:2002.11739 [hep-ph]}}.

\bibitem{Ema:2020evi}
Y.~Ema, K.~Mukaida, and J.~van~de Vis, ``{Renormalization group equations of Higgs-R$^{2}$ inflation},'' \href{http://dx.doi.org/10.1007/JHEP02(2021)109}{{\em JHEP} {\bfseries 02} (2021) 109}, \href{http://arxiv.org/abs/2008.01096}{{\ttfamily arXiv:2008.01096 [hep-ph]}}.

\bibitem{Gorbunov:2018llf}
D.~Gorbunov and A.~Tokareva, ``{Scalaron the healer: removing the strong-coupling in the Higgs- and Higgs-dilaton inflations},'' \href{http://dx.doi.org/10.1016/j.physletb.2018.11.015}{{\em Phys. Lett. B} {\bfseries 788} (2019) 37--41}, \href{http://arxiv.org/abs/1807.02392}{{\ttfamily arXiv:1807.02392 [hep-ph]}}.

\bibitem{He:2018mgb}
M.~He, R.~Jinno, K.~Kamada, S.~C. Park, A.~A. Starobinsky, and J.~Yokoyama, ``{On the violent preheating in the mixed Higgs-$R^2$ inflationary model},'' \href{http://dx.doi.org/10.1016/j.physletb.2019.02.008}{{\em Phys. Lett. B} {\bfseries 791} (2019) 36--42}, \href{http://arxiv.org/abs/1812.10099}{{\ttfamily arXiv:1812.10099 [hep-ph]}}.

\bibitem{Wang:2017fuy}
Y.-C. Wang and T.~Wang, ``{Primordial perturbations generated by Higgs field and $R^2$ operator},'' \href{http://dx.doi.org/10.1103/PhysRevD.96.123506}{{\em Phys. Rev. D} {\bfseries 96} no.~12, (2017) 123506}, \href{http://arxiv.org/abs/1701.06636}{{\ttfamily arXiv:1701.06636 [gr-qc]}}.

\bibitem{He:2018gyf}
M.~He, A.~A. Starobinsky, and J.~Yokoyama, ``{Inflation in the mixed Higgs-$R^2$ model},'' \href{http://dx.doi.org/10.1088/1475-7516/2018/05/064}{{\em JCAP} {\bfseries 05} (2018) 064}, \href{http://arxiv.org/abs/1804.00409}{{\ttfamily arXiv:1804.00409 [astro-ph.CO]}}.

\bibitem{Gundhi:2018wyz}
A.~Gundhi and C.~F. Steinwachs, ``{Scalaron-Higgs inflation},'' \href{http://dx.doi.org/10.1016/j.nuclphysb.2020.114989}{{\em Nucl. Phys. B} {\bfseries 954} (2020) 114989}, \href{http://arxiv.org/abs/1810.10546}{{\ttfamily arXiv:1810.10546 [hep-th]}}.

\bibitem{Enckell:2018uic}
V.-M. Enckell, K.~Enqvist, S.~Rasanen, and L.-P. Wahlman, ``{Higgs-$R^2$ inflation - full slow-roll study at tree-level},'' \href{http://dx.doi.org/10.1088/1475-7516/2020/01/041}{{\em JCAP} {\bfseries 01} (2020) 041}, \href{http://arxiv.org/abs/1812.08754}{{\ttfamily arXiv:1812.08754 [astro-ph.CO]}}.

\bibitem{Bezrukov:2019ylq}
F.~Bezrukov, D.~Gorbunov, C.~Shepherd, and A.~Tokareva, ``{Some like it hot: $R^2$ heals Higgs inflation, but does not cool it},'' \href{http://dx.doi.org/10.1016/j.physletb.2019.06.064}{{\em Phys. Lett. B} {\bfseries 795} (2019) 657--665}, \href{http://arxiv.org/abs/1904.04737}{{\ttfamily arXiv:1904.04737 [hep-ph]}}.

\bibitem{He:2020ivk}
M.~He, R.~Jinno, K.~Kamada, A.~A. Starobinsky, and J.~Yokoyama, ``{Occurrence of tachyonic preheating in the mixed Higgs-R$^2$ model},'' \href{http://dx.doi.org/10.1088/1475-7516/2021/01/066}{{\em JCAP} {\bfseries 01} (2021) 066}, \href{http://arxiv.org/abs/2007.10369}{{\ttfamily arXiv:2007.10369 [hep-ph]}}.

\bibitem{Bezrukov:2020txg}
F.~Bezrukov and C.~Shepherd, ``{A heatwave affair: mixed Higgs-$R^2$ preheating on the lattice},'' \href{http://dx.doi.org/10.1088/1475-7516/2020/12/028}{{\em JCAP} {\bfseries 12} (2020) 028}, \href{http://arxiv.org/abs/2007.10978}{{\ttfamily arXiv:2007.10978 [hep-ph]}}.

\bibitem{He:2020qcb}
M.~He, ``{Perturbative Reheating in the Mixed Higgs-$R^2$ Model},'' \href{http://dx.doi.org/10.1088/1475-7516/2021/05/021}{{\em JCAP} {\bfseries 05} (2021) 021}, \href{http://arxiv.org/abs/2010.11717}{{\ttfamily arXiv:2010.11717 [hep-ph]}}.

\bibitem{Aoki:2022dzd}
S.~Aoki, H.~M. Lee, A.~G. Menkara, and K.~Yamashita, ``{Reheating and dark matter freeze-in in the Higgs-R$^{2}$ inflation model},'' \href{http://dx.doi.org/10.1007/JHEP05(2022)121}{{\em JHEP} {\bfseries 05} (2022) 121}, \href{http://arxiv.org/abs/2202.13063}{{\ttfamily arXiv:2202.13063 [hep-ph]}}.

\bibitem{Cheong:2019vzl}
D.~Y. Cheong, S.~M. Lee, and S.~C. Park, ``{Primordial black holes in Higgs-$R^2$ inflation as the whole of dark matter},'' \href{http://dx.doi.org/10.1088/1475-7516/2021/01/032}{{\em JCAP} {\bfseries 01} (2021) 032}, \href{http://arxiv.org/abs/1912.12032}{{\ttfamily arXiv:1912.12032 [hep-ph]}}.

\bibitem{Cheong:2022gfc}
D.~Y. Cheong, K.~Kohri, and S.~C. Park, ``{The inflaton that could: primordial black holes and second order gravitational waves from tachyonic instability induced in Higgs-R $^{2}$ inflation},'' \href{http://dx.doi.org/10.1088/1475-7516/2022/10/015}{{\em JCAP} {\bfseries 10} (2022) 015}, \href{http://arxiv.org/abs/2205.14813}{{\ttfamily arXiv:2205.14813 [hep-ph]}}.

\bibitem{Durrer:2022emo}
R.~Durrer, O.~Sobol, and S.~Vilchinskii, ``{Magnetogenesis in Higgs-Starobinsky inflation},'' \href{http://dx.doi.org/10.1103/PhysRevD.106.123520}{{\em Phys. Rev. D} {\bfseries 106} no.~12, (2022) 123520}, \href{http://arxiv.org/abs/2207.05030}{{\ttfamily arXiv:2207.05030 [gr-qc]}}.

\bibitem{Cheong:2021vdb}
D.~Y. Cheong, S.~M. Lee, and S.~C. Park, ``{Progress in Higgs inflation},'' \href{http://dx.doi.org/10.1007/s40042-021-00086-2}{{\em J. Korean Phys. Soc.} {\bfseries 78} no.~10, (2021) 897--906}, \href{http://arxiv.org/abs/2103.00177}{{\ttfamily arXiv:2103.00177 [hep-ph]}}.

\bibitem{Canko:2019mud}
D.~D. Canko, I.~D. Gialamas, and G.~P. Kodaxis, ``{A simple $F(\mathcal{R},\phi )$ deformation of Starobinsky inflationary model},'' \href{http://dx.doi.org/10.1140/epjc/s10052-020-8025-4}{{\em Eur. Phys. J. C} {\bfseries 80} no.~5, (2020) 458}, \href{http://arxiv.org/abs/1901.06296}{{\ttfamily arXiv:1901.06296 [hep-th]}}.

\bibitem{Lee:2021dgi}
H.~M. Lee and A.~G. Menkara, ``{Cosmology of linear Higgs-sigma models with conformal invariance},'' \href{http://dx.doi.org/10.1007/JHEP09(2021)018}{{\em JHEP} {\bfseries 09} (2021) 018}, \href{http://arxiv.org/abs/2104.10390}{{\ttfamily arXiv:2104.10390 [hep-ph]}}.

\bibitem{Bauer:2010jg}
F.~Bauer and D.~A. Demir, ``{Higgs-Palatini Inflation and Unitarity},'' \href{http://dx.doi.org/10.1016/j.physletb.2011.03.042}{{\em Phys. Lett. B} {\bfseries 698} (2011) 425--429}, \href{http://arxiv.org/abs/1012.2900}{{\ttfamily arXiv:1012.2900 [hep-ph]}}.

\bibitem{Rasanen:2018ihz}
S.~Rasanen, ``{Higgs inflation in the Palatini formulation with kinetic terms for the metric},'' \href{http://dx.doi.org/10.21105/astro.1811.09514}{{\em Open J. Astrophys.} {\bfseries 2} no.~1, (2019) 1}, \href{http://arxiv.org/abs/1811.09514}{{\ttfamily arXiv:1811.09514 [gr-qc]}}.

\bibitem{Rubio:2019ypq}
J.~Rubio and E.~S. Tomberg, ``{Preheating in Palatini Higgs inflation},'' \href{http://dx.doi.org/10.1088/1475-7516/2019/04/021}{{\em JCAP} {\bfseries 04} (2019) 021}, \href{http://arxiv.org/abs/1902.10148}{{\ttfamily arXiv:1902.10148 [hep-ph]}}.

\bibitem{Gialamas:2019nly}
I.~D. Gialamas and A.~B. Lahanas, ``{Reheating in $R^2$ Palatini inflationary models},'' \href{http://dx.doi.org/10.1103/PhysRevD.101.084007}{{\em Phys. Rev. D} {\bfseries 101} no.~8, (2020) 084007}, \href{http://arxiv.org/abs/1911.11513}{{\ttfamily arXiv:1911.11513 [gr-qc]}}.

\bibitem{Cheong:2021kyc}
D.~Y. Cheong, S.~M. Lee, and S.~C. Park, ``{Reheating in models with non-minimal coupling in metric and~Palatini formalisms},'' \href{http://dx.doi.org/10.1088/1475-7516/2022/02/029}{{\em JCAP} {\bfseries 02} no.~02, (2022) 029}, \href{http://arxiv.org/abs/2111.00825}{{\ttfamily arXiv:2111.00825 [hep-ph]}}.

\bibitem{Lahanas:2022mng}
A.~B. Lahanas, ``{Issues in Palatini R2 inflation: Bounds on the reheating temperature},'' \href{http://dx.doi.org/10.1103/PhysRevD.106.123530}{{\em Phys. Rev. D} {\bfseries 106} no.~12, (2022) 123530}, \href{http://arxiv.org/abs/2210.00837}{{\ttfamily arXiv:2210.00837 [gr-qc]}}.

\bibitem{Enckell:2018hmo}
V.-M. Enckell, K.~Enqvist, S.~Rasanen, and L.-P. Wahlman, ``{Inflation with $R^2$ term in the Palatini formalism},'' \href{http://dx.doi.org/10.1088/1475-7516/2019/02/022}{{\em JCAP} {\bfseries 02} (2019) 022}, \href{http://arxiv.org/abs/1810.05536}{{\ttfamily arXiv:1810.05536 [gr-qc]}}.

\bibitem{Antoniadis:2018ywb}
I.~Antoniadis, A.~Karam, A.~Lykkas, and K.~Tamvakis, ``{Palatini inflation in models with an $R^2$ term},'' \href{http://dx.doi.org/10.1088/1475-7516/2018/11/028}{{\em JCAP} {\bfseries 11} (2018) 028}, \href{http://arxiv.org/abs/1810.10418}{{\ttfamily arXiv:1810.10418 [gr-qc]}}.

\bibitem{Jinno:2019und}
R.~Jinno, M.~Kubota, K.-y. Oda, and S.~C. Park, ``{Higgs inflation in metric and Palatini formalisms: Required suppression of higher dimensional operators},'' \href{http://dx.doi.org/10.1088/1475-7516/2020/03/063}{{\em JCAP} {\bfseries 03} (2020) 063}, \href{http://arxiv.org/abs/1904.05699}{{\ttfamily arXiv:1904.05699 [hep-ph]}}.

\bibitem{Mikura:2021clt}
Y.~Mikura and Y.~Tada, ``{On UV-completion of Palatini-Higgs inflation},'' \href{http://dx.doi.org/10.1088/1475-7516/2022/05/035}{{\em JCAP} {\bfseries 05} no.~05, (2022) 035}, \href{http://arxiv.org/abs/2110.03925}{{\ttfamily arXiv:2110.03925 [hep-ph]}}.

\bibitem{He:2022xef}
M.~He, Y.~Mikura, and Y.~Tada, ``{Hybrid metric-Palatini Higgs inflation},'' \href{http://dx.doi.org/10.1088/1475-7516/2023/05/047}{{\em JCAP} {\bfseries 05} (2023) 047}, \href{http://arxiv.org/abs/2209.11051}{{\ttfamily arXiv:2209.11051 [hep-th]}}.

\bibitem{Shaposhnikov:2020gts}
M.~Shaposhnikov, A.~Shkerin, I.~Timiryasov, and S.~Zell, ``{Higgs inflation in Einstein-Cartan gravity},'' \href{http://dx.doi.org/10.1088/1475-7516/2021/10/E01}{{\em JCAP} {\bfseries 02} (2021) 008}, \href{http://arxiv.org/abs/2007.14978}{{\ttfamily arXiv:2007.14978 [hep-ph]}}. [Erratum: JCAP 10, E01 (2021)].

\bibitem{Piani:2023aof}
M.~Piani and J.~Rubio, ``{Preheating in Einstein-Cartan Higgs Inflation: Oscillon formation},'' \href{http://arxiv.org/abs/2304.13056}{{\ttfamily arXiv:2304.13056 [hep-ph]}}.

\bibitem{Nieh:1981ww}
H.~T. Nieh and M.~L. Yan, ``{An Identity in Riemann-cartan Geometry},'' \href{http://dx.doi.org/10.1063/1.525379}{{\em J. Math. Phys.} {\bfseries 23} (1982) 373}.

\bibitem{Carroll:2004st}
S.~M. Carroll, {\em {Spacetime and Geometry}: {An Introduction to General Relativity}}.
\newblock Cambridge University Press, 7, 2019.

\bibitem{Sotiriou:2006qn}
T.~P. Sotiriou and S.~Liberati, ``{Metric-affine f(R) theories of gravity},'' \href{http://dx.doi.org/10.1016/j.aop.2006.06.002}{{\em Annals Phys.} {\bfseries 322} (2007) 935--966}, \href{http://arxiv.org/abs/gr-qc/0604006}{{\ttfamily arXiv:gr-qc/0604006}}.

\bibitem{Sotiriou:2008rp}
T.~P. Sotiriou and V.~Faraoni, ``{f(R) Theories Of Gravity},'' \href{http://dx.doi.org/10.1103/RevModPhys.82.451}{{\em Rev. Mod. Phys.} {\bfseries 82} (2010) 451--497}, \href{http://arxiv.org/abs/0805.1726}{{\ttfamily arXiv:0805.1726 [gr-qc]}}.

\bibitem{Olmo:2011uz}
G.~J. Olmo, ``{Palatini Approach to Modified Gravity: f(R) Theories and Beyond},'' \href{http://dx.doi.org/10.1142/S0218271811018925}{{\em Int. J. Mod. Phys. D} {\bfseries 20} (2011) 413--462}, \href{http://arxiv.org/abs/1101.3864}{{\ttfamily arXiv:1101.3864 [gr-qc]}}.

\bibitem{Hehl:1994ue}
F.~W. Hehl, J.~D. McCrea, E.~W. Mielke, and Y.~Ne'eman, ``{Metric affine gauge theory of gravity: Field equations, Noether identities, world spinors, and breaking of dilation invariance},'' \href{http://dx.doi.org/10.1016/0370-1573(94)00111-F}{{\em Phys. Rept.} {\bfseries 258} (1995) 1--171}, \href{http://arxiv.org/abs/gr-qc/9402012}{{\ttfamily arXiv:gr-qc/9402012}}.

\bibitem{Julia:1998ys}
B.~Julia and S.~Silva, ``{Currents and superpotentials in classical gauge invariant theories. 1. Local results with applications to perfect fluids and general relativity},'' \href{http://dx.doi.org/10.1088/0264-9381/15/8/006}{{\em Class. Quant. Grav.} {\bfseries 15} (1998) 2173--2215}, \href{http://arxiv.org/abs/gr-qc/9804029}{{\ttfamily arXiv:gr-qc/9804029}}.

\bibitem{Dadhich:2012htv}
N.~Dadhich and J.~M. Pons, ``{On the equivalence of the Einstein-Hilbert and the Einstein-Palatini formulations of general relativity for an arbitrary connection},'' \href{http://dx.doi.org/10.1007/s10714-012-1393-9}{{\em Gen. Rel. Grav.} {\bfseries 44} (2012) 2337--2352}, \href{http://arxiv.org/abs/1010.0869}{{\ttfamily arXiv:1010.0869 [gr-qc]}}.

\bibitem{Shapiro:2014kma}
I.~L. Shapiro and P.~M. Teixeira, ``{Quantum Einstein-Cartan theory with the Holst term},'' \href{http://dx.doi.org/10.1088/0264-9381/31/18/185002}{{\em Class. Quant. Grav.} {\bfseries 31} (2014) 185002}, \href{http://arxiv.org/abs/1402.4854}{{\ttfamily arXiv:1402.4854 [hep-th]}}.

\bibitem{Alonso:2015fsp}
R.~Alonso, E.~E. Jenkins, and A.~V. Manohar, ``{A Geometric Formulation of Higgs Effective Field Theory: Measuring the Curvature of Scalar Field Space},'' \href{http://dx.doi.org/10.1016/j.physletb.2016.01.041}{{\em Phys. Lett. B} {\bfseries 754} (2016) 335--342}, \href{http://arxiv.org/abs/1511.00724}{{\ttfamily arXiv:1511.00724 [hep-ph]}}.

\bibitem{Alonso:2016oah}
R.~Alonso, E.~E. Jenkins, and A.~V. Manohar, ``{Geometry of the Scalar Sector},'' \href{http://dx.doi.org/10.1007/JHEP08(2016)101}{{\em JHEP} {\bfseries 08} (2016) 101}, \href{http://arxiv.org/abs/1605.03602}{{\ttfamily arXiv:1605.03602 [hep-ph]}}.

\bibitem{Nagai:2019tgi}
R.~Nagai, M.~Tanabashi, K.~Tsumura, and Y.~Uchida, ``{Symmetry and geometry in a generalized Higgs effective field theory: Finiteness of oblique corrections versus perturbative unitarity},'' \href{http://dx.doi.org/10.1103/PhysRevD.100.075020}{{\em Phys. Rev. D} {\bfseries 100} no.~7, (2019) 075020}, \href{http://arxiv.org/abs/1904.07618}{{\ttfamily arXiv:1904.07618 [hep-ph]}}.

\bibitem{Cohen:2021ucp}
T.~Cohen, N.~Craig, X.~Lu, and D.~Sutherland, ``{Unitarity violation and the geometry of Higgs EFTs},'' \href{http://dx.doi.org/10.1007/JHEP12(2021)003}{{\em JHEP} {\bfseries 12} (2021) 003}, \href{http://arxiv.org/abs/2108.03240}{{\ttfamily arXiv:2108.03240 [hep-ph]}}.

\bibitem{Cheung:2021yog}
C.~Cheung, A.~Helset, and J.~Parra-Martinez, ``{Geometric soft theorems},'' \href{http://dx.doi.org/10.1007/JHEP04(2022)011}{{\em JHEP} {\bfseries 04} (2022) 011}, \href{http://arxiv.org/abs/2111.03045}{{\ttfamily arXiv:2111.03045 [hep-th]}}.

\bibitem{Helset:2022tlf}
A.~Helset, E.~E. Jenkins, and A.~V. Manohar, ``{Geometry in scattering amplitudes},'' \href{http://dx.doi.org/10.1103/PhysRevD.106.116018}{{\em Phys. Rev. D} {\bfseries 106} no.~11, (2022) 116018}, \href{http://arxiv.org/abs/2210.08000}{{\ttfamily arXiv:2210.08000 [hep-ph]}}.

\bibitem{Hill:2020oaj}
C.~T. Hill and G.~G. Ross, ``{Gravitational Contact Interactions and the Physical Equivalence of Weyl Transformations in Effective Field Theory},'' \href{http://dx.doi.org/10.1103/PhysRevD.102.125014}{{\em Phys. Rev. D} {\bfseries 102} (2020) 125014}, \href{http://arxiv.org/abs/2009.14782}{{\ttfamily arXiv:2009.14782 [gr-qc]}}.

\bibitem{Karananas:2022byw}
G.~K. Karananas, M.~Shaposhnikov, and S.~Zell, ``{Field redefinitions, perturbative unitarity and Higgs inflation},'' \href{http://dx.doi.org/10.1007/JHEP06(2022)132}{{\em JHEP} {\bfseries 06} (2022) 132}, \href{http://arxiv.org/abs/2203.09534}{{\ttfamily arXiv:2203.09534 [hep-ph]}}.

\bibitem{Mukohyama:2020lsu}
S.~Mukohyama and R.~Namba, ``{Partial UV Completion of $P(X)$ from a Curved Field Space},'' \href{http://dx.doi.org/10.1088/1475-7516/2021/02/001}{{\em JCAP} {\bfseries 02} (2021) 001}, \href{http://arxiv.org/abs/2010.09184}{{\ttfamily arXiv:2010.09184 [hep-th]}}.

\bibitem{Bardeen:1976zh}
W.~A. Bardeen, B.~W. Lee, and R.~E. Shrock, ``{Phase Transition in the Nonlinear $\sigma$ Model in 2 + $\epsilon$ Dimensional Continuum},'' \href{http://dx.doi.org/10.1103/PhysRevD.14.985}{{\em Phys. Rev. D} {\bfseries 14} (1976) 985}.

\bibitem{Brezin:1976ap}
E.~Brezin, J.~Zinn-Justin, and J.~C. Le~Guillou, ``{Renormalization of the Nonlinear Sigma Model in (Two + Epsilon) Dimension},'' \href{http://dx.doi.org/10.1103/PhysRevD.14.2615}{{\em Phys. Rev. D} {\bfseries 14} (1976) 2615}.

\bibitem{Dobado:1995qy}
A.~Dobado, M.~J. Herrero, J.~R. Pelaez, E.~Ruiz~Morales, and M.~T. Urdiales, ``{Learning about the strongly interacting symmetry breaking sector at LHC},'' \href{http://dx.doi.org/10.1016/0370-2693(95)00431-J}{{\em Phys. Lett. B} {\bfseries 352} (1995) 400--410}, \href{http://arxiv.org/abs/hep-ph/9502309}{{\ttfamily arXiv:hep-ph/9502309}}.

\bibitem{Dobado:1999xb}
A.~Dobado, M.~J. Herrero, J.~R. Pelaez, and E.~Ruiz~Morales, ``{CERN LHC sensitivity to the resonance spectrum of a minimal strongly interacting electroweak symmetry breaking sector},'' \href{http://dx.doi.org/10.1103/PhysRevD.62.055011}{{\em Phys. Rev. D} {\bfseries 62} (2000) 055011}, \href{http://arxiv.org/abs/hep-ph/9912224}{{\ttfamily arXiv:hep-ph/9912224}}.

\bibitem{DAdda:1978vbw}
A.~D'Adda, M.~Luscher, and P.~Di~Vecchia, ``{A 1/n Expandable Series of Nonlinear Sigma Models with Instantons},'' \href{http://dx.doi.org/10.1016/0550-3213(78)90432-7}{{\em Nucl. Phys. B} {\bfseries 146} (1978) 63--76}.

\bibitem{DAdda:1978dle}
A.~D'Adda, P.~Di~Vecchia, and M.~Luscher, ``{Confinement and Chiral Symmetry Breaking in CP**n-1 Models with Quarks},'' \href{http://dx.doi.org/10.1016/0550-3213(79)90083-X}{{\em Nucl. Phys. B} {\bfseries 152} (1979) 125--144}.

\bibitem{Witten:1978bc}
E.~Witten, ``{Instantons, the Quark Model, and the 1/n Expansion},'' \href{http://dx.doi.org/10.1016/0550-3213(79)90243-8}{{\em Nucl. Phys. B} {\bfseries 149} (1979) 285--320}.

\bibitem{Arefeva:1980ms}
I.~Y. Arefeva and S.~I. Azakov, ``{RENORMALIZATION AND PHASE TRANSITION IN THE QUANTUM CP**(n-1) MODEL (D = 2, 3)},'' \href{http://dx.doi.org/10.1016/0550-3213(80)90266-7}{{\em Nucl. Phys. B} {\bfseries 162} (1980) 298--310}.

\bibitem{Fujiwara:1984mp}
T.~Fujiwara, T.~Kugo, H.~Terao, S.~Uehara, and K.~Yamawaki, ``{Nonabelian Anomaly and Vector Mesons as Dynamical Gauge Bosons of Hidden Local Symmetries},'' \href{http://dx.doi.org/10.1143/PTP.73.926}{{\em Prog. Theor. Phys.} {\bfseries 73} (1985) 926}.

\bibitem{Bando:1984ej}
M.~Bando, T.~Kugo, S.~Uehara, K.~Yamawaki, and T.~Yanagida, ``{Is rho Meson a Dynamical Gauge Boson of Hidden Local Symmetry?},'' \href{http://dx.doi.org/10.1103/PhysRevLett.54.1215}{{\em Phys. Rev. Lett.} {\bfseries 54} (1985) 1215}.

\bibitem{Bando:1987br}
M.~Bando, T.~Kugo, and K.~Yamawaki, ``{Nonlinear Realization and Hidden Local Symmetries},'' \href{http://dx.doi.org/10.1016/0370-1573(88)90019-1}{{\em Phys. Rept.} {\bfseries 164} (1988) 217--314}.

\bibitem{Nambu:1961tp}
Y.~Nambu and G.~Jona-Lasinio, ``{Dynamical Model of Elementary Particles Based on an Analogy with Superconductivity. 1.},'' \href{http://dx.doi.org/10.1103/PhysRev.122.345}{{\em Phys. Rev.} {\bfseries 122} (1961) 345--358}.

\bibitem{Nambu:1961fr}
Y.~Nambu and G.~Jona-Lasinio, ``{Dynamical model of elementary particles based on an analogy with superconductivity. II.},'' \href{http://dx.doi.org/10.1103/PhysRev.124.246}{{\em Phys. Rev.} {\bfseries 124} (1961) 246--254}.

\bibitem{Gross:1974jv}
D.~J. Gross and A.~Neveu, ``{Dynamical Symmetry Breaking in Asymptotically Free Field Theories},'' \href{http://dx.doi.org/10.1103/PhysRevD.10.3235}{{\em Phys. Rev. D} {\bfseries 10} (1974) 3235}.

\bibitem{Dvali:2007hz}
G.~Dvali, ``{Black Holes and Large N Species Solution to the Hierarchy Problem},'' \href{http://dx.doi.org/10.1002/prop.201000009}{{\em Fortsch. Phys.} {\bfseries 58} (2010) 528--536}, \href{http://arxiv.org/abs/0706.2050}{{\ttfamily arXiv:0706.2050 [hep-th]}}.

\bibitem{Han:2004wt}
T.~Han and S.~Willenbrock, ``{Scale of quantum gravity},'' \href{http://dx.doi.org/10.1016/j.physletb.2005.04.040}{{\em Phys. Lett. B} {\bfseries 616} (2005) 215--220}, \href{http://arxiv.org/abs/hep-ph/0404182}{{\ttfamily arXiv:hep-ph/0404182}}.

\bibitem{Ema:2021xhq}
Y.~Ema, R.~Jinno, K.~Nakayama, and J.~van~de Vis, ``{Preheating from target space curvature and unitarity violation: Analysis in field space},'' \href{http://dx.doi.org/10.1103/PhysRevD.103.103536}{{\em Phys. Rev. D} {\bfseries 103} no.~10, (2021) 103536}, \href{http://arxiv.org/abs/2102.12501}{{\ttfamily arXiv:2102.12501 [hep-ph]}}.

\bibitem{Dohse:2018vqo}
M.~Dohse, ``{TikZ-FeynHand: Basic User Guide},'' \href{http://arxiv.org/abs/1802.00689}{{\ttfamily arXiv:1802.00689 [cs.OH]}}.

\bibitem{Ellis:2016jkw}
J.~Ellis, ``{TikZ-Feynman: Feynman diagrams with TikZ},'' \href{http://dx.doi.org/10.1016/j.cpc.2016.08.019}{{\em Comput. Phys. Commun.} {\bfseries 210} (2017) 103--123}, \href{http://arxiv.org/abs/1601.05437}{{\ttfamily arXiv:1601.05437 [hep-ph]}}.

\end{thebibliography}\endgroup
%%%%%%%%%%%%%%%%%%%%%%%%%%%%%%%%%%%%%%%%%%%%%%%%%%

%%%%%%%%%%%%%%%%%%%%%%%%%%%%%%%%%%%%%%%%%%%%%%%%%%
\end{document}